\documentclass[a4paper,fleqn]{cas-sc}

\usepackage[numbers]{natbib}
\usepackage{amsmath,amssymb,amsfonts}
\usepackage{algorithm}
\usepackage{graphicx}
\usepackage{textcomp}
\usepackage{xcolor}
\usepackage{booktabs}
\usepackage{multirow}
\usepackage{hyperref}
\usepackage{algpseudocode}
\usepackage{comment}
\usepackage{stfloats}

\def\tsc#1{\csdef{#1}{\textsc{\lowercase{#1}}\xspace}}
\tsc{WGM}
\tsc{QE}
\tsc{EP}
\tsc{PMS}
\tsc{BEC}
\tsc{DE}

\begin{document}
\let\WriteBookmarks\relax
\def\floatpagepagefraction{1}
\def\textpagefraction{.001}

\shorttitle{CA-QIRL}


\title [mode = title]{Communication-Aware Quantum-Inspired Reinforcement Learning for Cyber-Resilient V2X Intrusion Detection and Mitigation}




\author[1]{Sajid Anwer}
\credit{Conceptualization, Methodology, Software, Funding acquisition, Writing -- original draft, Writing -- review and editing}

\author[2]{Rohan Farooq}
\credit{Conceptualization, Methodology, Software}

\author[2]{Anwar Shah}
\cormark[1]
\ead{anwar.shah@nu.edu.pk}
\credit{Conceptualization, Methodology, Software, Project administration}

\author[1]{Tallha Akram}
\credit{Conceptualization, Methodology, Writing -- original draft, Writing -- review and editing}


\affiliation[1]{
organization={Department of Software Engineering, College of Computer Engineering and Sciences, Prince Sattam Bin Abdulaziz University},
country={Saudi Arabia}
}

\affiliation[2]{
organization={Department of Computer Science, National University of Computer and Emerging Sciences},
country={Pakistan}
}


\cortext[cor1]{Corresponding author}

\nonumnote{
Author emails: Sajid Anwer, \texttt{s.anwer@psau.edu.sa}; 
Rohan Farooq, \texttt{rohan.farooq@nu.edu.pk}; 
Tallha Akram, \texttt{t.akram@psau.edu.sa}.
}
    
\begin{abstract}
Smart cities increasingly depend on dense edge, IoT, and vehicular networks to deliver critical urban services, including traffic control, connected mobility, infrastructure monitoring, and energy management. In this ecosystem, the Internet of Vehicles (IoV) is central to intelligent transportation, enabling continuous communication among vehicles, roadside infrastructure, and cloud-edge platforms. This connectivity, however, also enlarges the attack surface and exposes smart city and vehicular systems to evolving cyber threats that can compromise safety, privacy, data integrity, and service continuity. Conventional static defenses are often inadequate because they cannot autonomously adapt to changing attack behaviors or multi-stage intrusion patterns. This paper proposes Communication Aware Quantum Inspired Reinforcement Learning (CA-QIRL) framework built on a lightweight deep Q-Network architecture for next-generation autonomous cyber defense. V2X defense is formulated as a communication-aware Markov Decision Process (MDP), with the agent observing intrusion, mobility, Road Side Unit (RSU), and communication metrics, followed by selecting mitigation actions. CA-QIRL combines quantum-inspired encoding, rotation exploration, interference reward, and cost function penalising false negatives, false positives, delay, packet loss, and RSU overload. Experimental evaluation on vehicular intrusion datasets (Car-Hacking, ROAD, VeReMi, CAN-MIRGU) and a mobility-aware V2X simulation demonstrates that the proposed framework achieves competitive detection accuracy of 97.89\% on CICIDS2017 and 80.31\% on CAN-MIRGU, while outperforming state-of-the-art ensemble methods in inference latency by factors of 69.2 times and 33.0 times, respectively. Moreover, the end-to-end delay and Channel Busy Ratio (CBR) are reduced by up to 95.7\% and 90\%, respectively, all while maintaining sub-100 $\mu$s inference latency across all datasets. Furthermore, statistical significance is demonstrated on ROAD (Recall: 0.985 vs. 0.951, $p < 0.01$) and VeReMi (Recall: 0.626 vs. 0.074, $p < 0.01$). These results suggest that Communication-Aware CA-QIRL is a practical cyber-resilient defense mechanism for next-generation V2X and Internet-of-Vehicles networks.

\end{abstract}




\begin{keywords}
\sep Smart Cities \sep Internet of Vehicles (IoV) \sep Cyber Defense \sep Reinforcement Learning \sep Quantum Computing 
\end{keywords}

\maketitle

\section{Introduction}

\noindent Smart cities is a new paradigm for urban management. It integrate a diverse network of IoT and edge devices to optimize critical services. These include traffic control, energy systems, and public safety infrastructure. The IoV extends this ecosystem through intelligent and connected transportation systems. Unlike static smart city infrastructure, vehicular networks introduce unique vulnerabilities. These are stemming from high node mobility, heterogeneous communication protocols, and ephemeral network topologies. These factors change as vehicles traverse different coverage zones. More critically, any defensive response in autonomous vehicle systems must operate within sub-millisecond safety windows. Therefore, this delayed threat detection directly translates to physical harm through compromised braking, steering, and collision avoidance systems. These dense and highly interconnected environments expose a wide attack surface to sophisticated cyber threats. Moreover, they face multi-stage attack campaigns that evolve dynamically over time. 

\noindent Traditional static defense systems are designed around fixed signatures and rule-based detection. Fundamentally, they cannot adapt to this evolving threat landscape. This limitation leaves critical autonomous vehicle safety systems and smart city infrastructure vulnerable to coordinated and context-aware attacks. Addressing this requires a defense mechanism that is simultaneously accurate, adaptive, and real-time capable. Additionally, it must be deployable on resource-constrained edge hardware.

\noindent Recent intrusion detection research has progressed through three broad directions, each advancing capability while introducing new limitations. Early machine learning based approaches applied classical algorithms to network traffic classification with promising results. Injadat et al.~\cite{injadat2021multi} proposed a multi-stage ML framework with information gain feature selection achieving 99.00\% accuracy, while  Alshammari et al.~\cite{alshammari2018classification} applied KNN and SVM to CAN bus data achieving 93\% and 96\% accuracy respectively, and Zhang et al.~\cite{zhang2019intrusion} combined genetic  algorithms with Deep Belief Networks on NSL-KDD achieving 98\% accuracy. Although these  methods demonstrated strong classification performance, they rely on static decision boundaries that cannot adapt to evolving attack distributions, and their inference latency typically exceeds the sub-millisecond safety threshold required for autonomous vehicle deployment~\cite{uddin2025scalable, survey2025ensemble}. Moreover, static supervised classifiers treat each network packet as an independent observation, ignoring the temporal correlations that characterize real-world multi-stage attack campaigns. Furthermore, standard loss functions fail to account for the asymmetric misclassification costs arising from the severely imbalanced traffic distributions encountered in IoV environments.

\noindent Deep learning approaches have addressed this limitation by learning hierarchical representations directly from raw traffic data. Nie et al.~\cite{nie2020data} proposed a CNN-based IDS for intelligent IoV roadside units, achieving an accuracy of 97.60\%, Kang and Kang~\cite{kang2016intrusion} developed a Deep  Neural Network for in-vehicle CAN security, reporting 97.80\%. In another study, Ashraf et al.~\cite{ashraf2021novel} proposed an LSTM autoencoder for intelligent transportation system anomaly detection, achieving an accuracy of 98-99\% on Car-Hacking by learning temporal patterns of normal traffic. Despite improved accuracy, deep learning models treat each packet as an independent observation, ignoring temporal correlations of multi-stage attack campaigns, and their inference cost grows with model depth, making real-time edge deployment increasingly challenging~\cite{dasari2025, secure2025dl}. 

\noindent Reinforcement learning approaches emerged to address the adaptivity gap of static classifiers by enabling agents to learn detection policies through interaction with the environment. In a formal model, Nguyen and Reddi~\cite{nguyen2018deep} represent network intrusion detection as a sequence of decisions. By using deep reinforcement learning, they show that systems change their behavior when threats evolve. Zhang et al.~\cite{zhang2024fair} proposed federated learning to detect anomalies in 6G networks, which shows that systems learn while they are online. However current reinforcement learning based methods for intrusion detection are not computationally effective, while achieving a stable state due to infrequent rewards. These methods do not function well due to high dimensional feature space.~\cite{survey2025ensemble, uddin2025scalable}. Reinforcement learning approaches have shown potential in adaptive intrusion detection but remain slow to converge and limited by reward sparsity in high-dimensional feature spaces. To date, no documented approach simultaneously satisfies competitive detection accuracy and sub-millisecond inference latency on standardized IoV datasets, a dual constraint that defines the central gap this work addresses.

\noindent Ensemble methods currently represent the state of the art in detection accuracy achieving 99.74\% and 100\% on  CICIDS2017 and UNSW-NB15, respectively  ~\cite{basepaper2025}. However, ensemble classifiers composed of multiple decision trees, each input sample to traverse all constituent models for prediction, making inference computationally intensive and fundamentally incompatible with the sub-millisecond safety threshold mandated for autonomous braking and steering interventions~\cite{uddin2025scalable, jha2025, survey2025ensemble}.These limitations cannot be completely interpreted by accuracy alone. 

\noindent In this work, we propose CA-QIRL that propose a accuracy-latency trade-off in a unified architecture. It embed quantum-inspired state encoding, interference-based training stabilization, and cost-sensitive reward augmentation into a lightweight Deep Q-Network. It achieves competitive detection accuracy, maintaining sub-50-microsecond inference latency, and suitable for real-time deployment on resource-constrained autonomous-vehicle edge hardware. 

\noindent The main contributions of this work are as follows. 
\begin{enumerate}
    \item Cyber-resilient reinforcement learning framework for V2X intrusion detection and mitigation that jointly considers traffic-level security features, vehicular mobility context, RSU/edge load, and communication quality indicators such as packet delivery ratio, end-to-end delay, beacon loss, and channel congestion.
    
    \item Markov Decision Process where the agent selects adaptive mitigation actions ($allow$, $alert$, $drop$, $isolate\_vehicle$, $reroute\_rsu$, $migrate\_task$, $reauthenticate$) to optimise both cybersecurity and vehicular communication reliability.
    
    \item Amplitude-phase state encoding, TD-error-guided rotation exploration, and temporal interference-based reward augmentation that improve temporal coherence without adding trainable parameters or inference-time overhead.
    
    \item Cost function that jointly penalises false negatives, false positives, communication delay, packet loss, and RSU overload, aligning intrusion detection with the operational requirements of safety-critical V2X environments.
    
    \item Validation on general intrusion datasets (CICIDS2017, UNSW-NB15), IoV-specific datasets (Car-Hacking, ROAD, VeReMi, CAN-MIRGU), and a mobility-aware V2X simulation, including detection metrics, communication metrics, and three-tier latency analysis.
\end{enumerate}

\noindent The remainder of this paper is organized as follows. In Section 2, the authors summarize previous research that uses traditional and modern methods for intrusion detection. To define the scope of the study, Sections 3 and 4 describe the problem, the threat model and the areas where an IoV is vulnerable to attack. For a technical explanation, Sections 5 and 6 discuss the details of the CA-QIRL framework. In Section 7, the results from experiments are provided, and Section 8 concludes the paper with directions for future work.

\section{Literature Review}
 
\noindent In this section, a detailed review of existing research on intrusion detection in IoV and smart city environments is covered using three progressively developed directions: traditional ML-based IDS, deep learning approaches, and reinforcement learning methods. This progression motivates the need for the adaptive, latency-aware, and communication aware framework.
 
\subsection{Machine Learning Based Intrusion Detection}

\noindent In the early stages of Intrusion Detection System research, authors used classical machine learning (ML) algorithms on network traffic data. To improve results, Injadat~et~al.~\cite{injadat2021multi} created a multi stage machine learning framework. This system uses information gain (IG) for the selection of features and includes oversampling. It is successful because it reaches a 99.00\% detection rate. By doing this the researchers showed that the engineering of features and the balancing of classes are necessary steps during the preprocessing stage. For another study, Alshammari~et~al.~\cite{alshammari2018classification} used K-Nearest Neighbors (KNN) besides Support Vector Machines (SVM) on Controller Area Network (CAN) bus data. Those methods are effective as they reach 93 \% and 96 \% accuracy but the authors did not provide data about the time the systems take to process information \cite{NETO2024101209}. And Zhang~et~al.~\cite{zhang2019intrusion} used a genetic algorithm with a Deep Belief Network (DBN) to detect attacks on Internet of Things devices using the NSL-KDD dataset. Their method is accurate at a level of 98 \%. To detect threats in autonomous vehicles, Alheeti and McDonald-Maier~\cite{alheeti2018intelligent} combined back propagation neural networks with fuzzy sets. On the KDD99 dataset, they reached a rate of 97.99 \%. While those ML methods are good at the task of classification, they are based on decision boundaries that do not change. Because of this lack of change, they are unable to adjust when the patterns of attacks evolve. If the systems are used, the time required for them to make a decision is usually more than the sub millisecond threshold that is necessary for safety in the Internet of Vehicles ~\cite{basepaper2025}.
 
\subsection{Deep Learning Based Intrusion Detection}
 
\noindent Deep learning improved upon classical ML by learning hierarchical representations from raw traffic. Nie~et~al.~\cite{nie2020data} proposed a CNN-based IDS for intelligent IoV roadside units, achieving 97.60\% by detecting abnormal link load patterns. Kang and Kang~\cite{kang2016intrusion} developed a DNN for in-vehicle CAN security using probability-based packet feature vectors, reporting 97.80\%. Ashraf~et~al.~\cite{ashraf2021novel} proposed an LSTM autoencoder for ITS anomaly detection, achieving 98--99\% on Car-Hacking by learning temporal patterns of normal traffic. Lokman~et~al.~\cite{lokman2018stacked} developed stacked sparse autoencoders for unsupervised CAN outlier detection, achieving 98\% F1 score. Recent hybrid DL systems have demonstrated improved IoT security: Wahab~et~al.~\cite{wahab2024sdn} proposed an SDN-based cognitive IDS, and Wahab~et~al.~\cite{wahab2025explainable} introduced a three-way neural network for explainable IoT intrusion detection. Ahmad~et~al.~\cite{ahmad2025survey} surveyed converging security paradigms including blockchain and digital twin-based defense frameworks. Despite improved accuracy, DL models treat each packet as an independent observation, ignoring the temporal correlations of multi-stage attack campaigns. Their inference cost also grows with model depth, making real-time edge deployment challenging \cite{R2025101666, GHOSH2024101129}.
 
\subsection{Reinforcement Learning and Adaptive Cyber Defense}
 
\noindent Reinforcement learning addresses the adaptivity gap of static classifiers by enabling agents to learn detection policies through environment interaction. Nguyen and Reddi~\cite{nguyen2018deep} formally modelled network intrusion detection as a sequential decision problem and demonstrated that DRL enables adaptive behavior under evolving threats. Al-Fuqaha~et~al.~\cite{al2020artificial} surveyed AI methods for IoV and identified RL as a promising direction for dynamic trust management and intrusion mitigation in vehicular networks. Zhang~et~al.~\cite{zhang2024fair} proposed a fair federated learning model for 6G network anomaly detection, demonstrating adaptive online learning. Ahmad et al. \cite{ahmad2025graphguard} demonstrate that a system learns to correct errors from malicious data injections in graph neural networks when it adjusts policies. By this result, the researchers indicate that defense mechanisms are useful when they change over time, but current methods for intrusion detection that use reinforcement learning encounter difficulties \cite{ZHAN2025101552, ZHU2024101192}. As an example, the learning process is slow because the feedback signals are infrequent. In addition, those systems fail to function when the data contains many different variables. On that account, the size of the feature space limits the technology. To date, no documented study shows that the systems process data in under 0.001 seconds. Because of this delay, the speed is not fast enough for the safety standards of cars that drive themselves.
 
\subsection{Ensemble Methods}
 
\noindent Ensemble methods currently achieve the highest reported accuracies. Ullah~et~al.~\cite{basepaper2025} proposed a hybrid stacking ensemble combining XGBoost, Random Forest, Decision Tree, and Gradient Boosting with IG feature selection, achieving 99.75\% and 100\%. However, the 100\% result applies SMOTE globally before splitting a data leakage condition and evaluates only a binary two-class. Under identical single-sample CPU conditions, their ensemble incurs 2,248\,\textmu{}s per flow, exceeding the autonomous vehicle safety threshold of 1\,ms by more than $2\times$.
The survey in Table~\ref{tab:comparison} confirms that no prior approach simultaneously satisfies competitive detection accuracy \emph{and} sub-millisecond inference latency on standardised IoV datasets. This dual constraint, together with the inability of static classifiers to adapt to evolving threats, motivates the proposed CA-QIRL framework.

\section{Problem Formulation}

\begin{figure}
    \centering
    \includegraphics[width=\columnwidth]{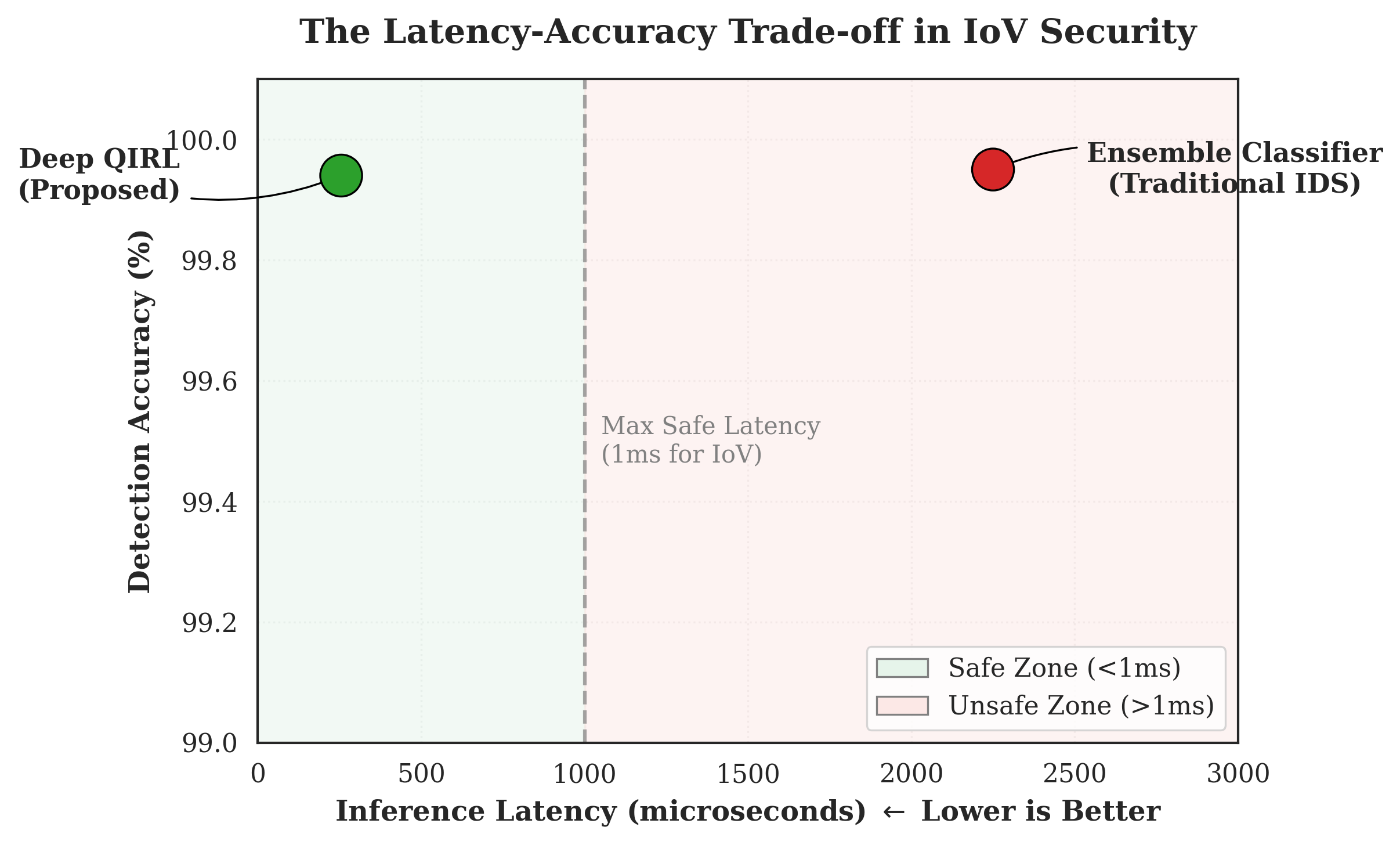}
    \caption{The critical trade-off between detection
    accuracy and inference latency in current IoV
    security frameworks. Existing ensemble methods
    occupy the high-accuracy but unsafe latency zone
    ($>\!1$ ms), while the proposed CA-QIRL framework
    simultaneously satisfies both constraints.}
    \label{fig:tradeoff}
\end{figure}

\noindent The Internet of Vehicles (IoV) grows rapidly and the attack surface for this system increases, respectively. Autonomous systems are at risk from these complex cyber threats that adapt to defenders strategies. In Fig.~\ref{fig:tradeoff}, there is a representation of recent IDS solutions existing gap that need to be resolved. Traditional ML ensemble methods like Random Forest, are accurate in detecting threats. However these methods are slow because the time they take to process data is more than $2$ ms/flow. To ensure safe autonomous braking and steering systems must act in less than 1 ms. While alternative methods are small and fast but not strong enough to identify threats against an adaptive threat.

\noindent Given network traffic represented as a sequence
of feature vectors $X = \{x_1, x_2, \dots, x_T\}$ where
each $x_t \in \mathbb{R}^d$ corresponds to a packet flow
with $d$ feature dimensions, the objective of any IDS is
to learn a mapping $f_\theta : X \rightarrow Y$ where
$Y = \{0, 1\}$ denotes benign and attack classes
respectively. Three structural gaps in current methodology
prevent this objective from being met under real-world
IoV deployment constraints. First, no existing approach
simultaneously satisfies the dual objective of minimizing
inference latency $\tau$ while maintaining detection
robustness above an acceptable threshold $\delta$:
\begin{equation} \label{eq:optimization}
    \min_{\theta} \; \tau(f_\theta) \quad
    \text{subject to} \quad \text{TSS}(f_\theta) \geq \delta
\end{equation}
where $\theta$ represents the model parameters and TSS
denotes the True Skill Statistic. Second, IoV datasets
exhibit complex and asymmetric class imbalance.
Standard loss functions bias the classifier toward
whichever class dominates, systematically elevating
either the False Negative Rate on attack-minority
datasets or the False Positive Rate on attack-majority
datasets:
\begin{equation} \label{eq:fnr}
    \text{FNR} = \frac{1}{N} \sum_{t=1}^{N}
    \mathbf{1}\bigl[f_\theta(x_t) = 0,\; y_t = 1\bigr]
\end{equation}
This demands a dataset-aware cost-sensitive formulation
with asymmetric penalties applied alongside training-only
SMOTE balancing to prevent data leakage:
\begin{equation} \label{eq:cost}
    J(\theta) = \frac{1}{N} \sum_{t=1}^{N}
    C\bigl(f_\theta(x_t),\, y_t\bigr)
\end{equation}
In this context $C(\cdot)$ is a cost matrix that follows the rule $C(0,1) > C(1,0) > C(0,0) = C(1,1) = 0$. By using those values, the system applies the largest numerical penalty to False Negatives, which are attacks that the system does not detect. To account for False Positives, the matrix applies a smaller numerical penalty when the system disrupts traffic that is not harmful. For instances where the classification is correct, the cost is zero. And as a third point, supervised learning methods process each packet $x_t$ as an independent data point. They ignore the
temporal correlation between consecutive flows:
\begin{equation} \label{eq:corr}
\text{Corr}(x_t, x_{t+1}) =
\frac{\text{Cov}(x_t,\, x_{t+1})}
{\sigma_{x_t}\, \sigma_{x_{t+1}}}
\neq 0
\end{equation}
There is a non zero correlation in actual attack campaigns that occur in multiple stages over time. Capturing this structure requires a Markov
Decision Process (MDP) formulation with state transition
$P(s_{t+1} \mid s_t, a_t)$, enabling the agent to
maximize the cumulative discounted reward:
\begin{equation} \label{eq:return}
    G_t = \sum_{k=0}^{\infty} \gamma^k\, r_{t+k}
\end{equation}

\begin{figure}
    \centering
    \includegraphics[width=1\linewidth]{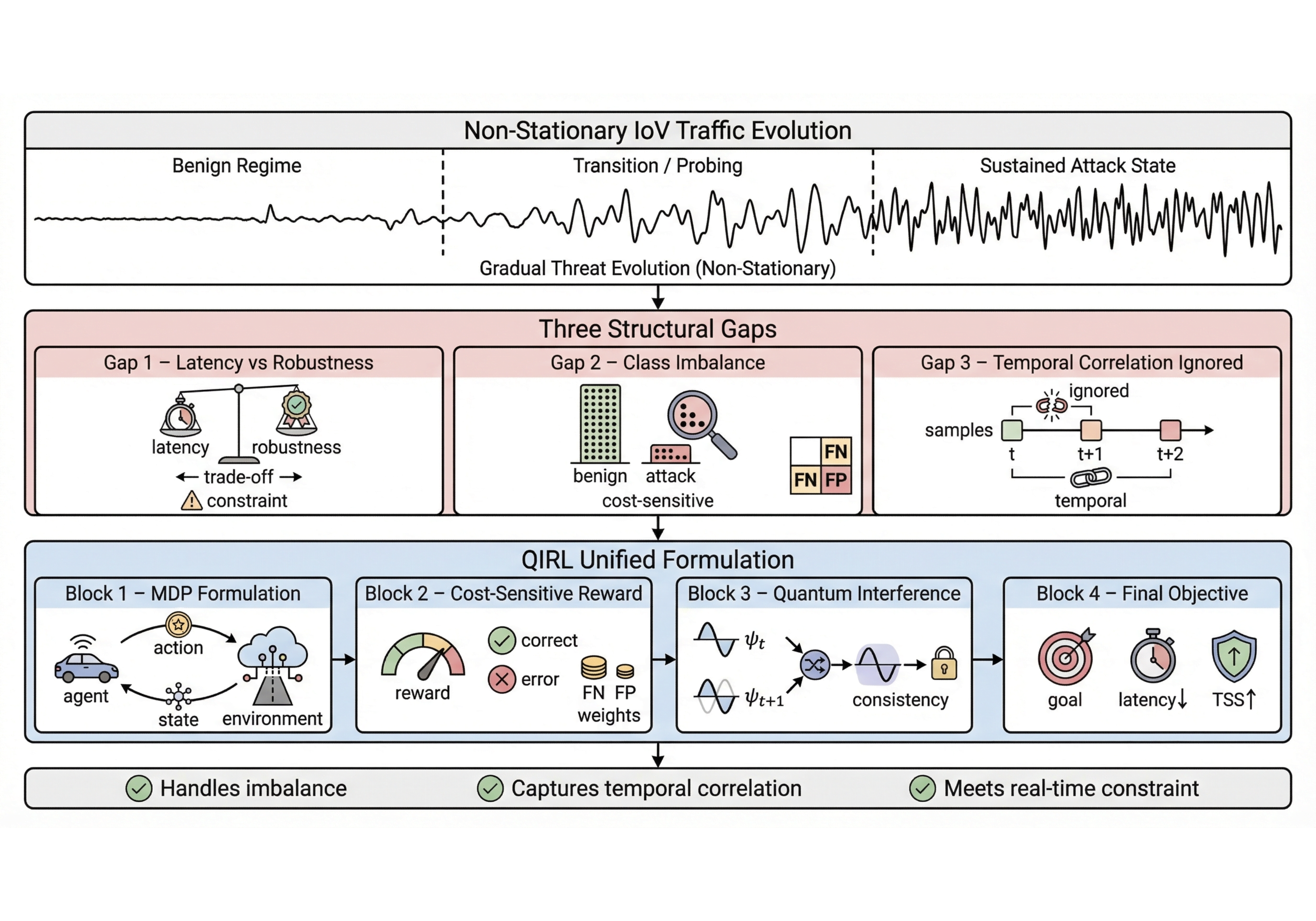}
    \caption{CA-QIRL problem formulation. IoV traffic exhibits non-stationary temporal behavior. Existing IDS methods fail due to latency–accuracy trade-offs, class imbalance, and ignored temporal correlation. CA-QIRL addresses these using a cost-sensitive MDP with quantum interference for efficient and robust detection.}
    \label{fig:problem_formulation}
\end{figure}

Figure \ref{fig:problem_formulation} summarizes these three structural gaps and the unified CA-QIRL formulation that addresses them.
where $\gamma \in [0,1]$ is the discount factor and
$r_{t+k}$ is the reward at step $t+k$, rather than
optimizing for greedy single-step accuracy. To address all three gaps within one system, we propose CA-QIRL: a Quantum Inspired Reinforcement Learning
architecture. It uses the cost function $J(\theta)$ from  Eq.~\ref{eq:cost}  as a reward signal with specific weights. By using those weights, the system manages situations where some categories of data are more frequent than others. As a second step we define the process of detection as a Markov Decision Process. This choice is made to represent how events relate to each other over time as shown in Eq.~\ref{eq:corr}. To improve the speed of the process, we use a small module based on quantum interference. Because of this module, the policy reaches the goals for time limits defined in eq.~\ref{eq:optimization} more quickly. When the system operates, the time required for a single decision is less than one millisecond. And the system remains as correct as other methods when it identifies threats in Internet of Vehicles settings, even if the settings have unequal amounts of data.

\section{Threat Model}

\noindent Effective cyber defense requires formally bounding the
adversary capabilities and attack surface before any
defense mechanism is evaluated. We define the adversary $A$. Five parameters define $A$
\begin{equation}
A = (K, O, G, P, T)
\end{equation}
In this equation $K$ describes how much information the adversary has about the network topology. It is represented by $O$ if the observation mode is passive sniffing or active probing. For the attack goal, $G$ indicates if the objective is disruption, exfiltration or hijacking. There is a probability $P \in [0,1]$ that the attack continues across time steps $t$. And $T$ specifies when the attack occurs over the sequence $\{t_1, t_2, \dots, t_n\}$.

\begin{table}
\centering
\caption{Vehicular Attack Types}
\label{tab:attacks}
\begin{tabular}{l|l}
\hline
\textbf{Attack} & \textbf{Vehicular Meaning} \\
\hline
DDoS on RSU & Edge/RSU overload from flooding \\
False Data Injection & Fake speed, position, acceleration, brake events \\
GPS Spoofing & Manipulated vehicle position/trajectory \\
Sybil Attack & Single vehicle claiming multiple identities \\
Replay Attack & Resending old safety messages \\
Jamming & Wireless channel disruption \\
Malicious RSU & Compromised infrastructure node \\
\hline
\end{tabular}
\end{table}

\noindent The IoV attack surface is distributed across four layers:
Vehicle-to-Infrastructure (V2I), Vehicle-to-Vehicle (V2V),
edge computing nodes, and cloud backend. For each layer
$i$, let $V_i$ denote the set of exploitable
vulnerabilities. The total attack surface is then:
\begin{equation}
    S = \sum_{i=1}^{4} |V_i|
\end{equation}
For any attack $\alpha \in V_i$, the adversary succeeds
when the IDS produces a False Negative, i.e., the
detection function $f(x_t) = 0$ when the true label
$y_t = 1$. The False Negative Rate across $N$ samples
is:
\begin{equation}
    \text{FNR} = \frac{1}{N} \sum_{t=1}^{N}
    \mathbf{1}\bigl[f(x_t) = 0 \;\wedge\; y_t = 1\bigr]
\end{equation}
The framework operates under a black-box assumption:
the adversary observes traffic flows $x_t$ but cannot
access or modify the reward signal $r_t$ or learned
policy parameters $\theta$ during inference. Physical
layer attacks (e.g., GPS spoofing) and adversarial model
perturbations are outside the current scope and are
deferred to future work. \ref{tab:attacks} summarizes the primary vehicular attack types considered within the threat model. Attacks within scope are modeled
as non-stationary processes where the probability of an
attack at step $t$ depends on prior state:
\begin{equation}
    P(\alpha_t \mid \alpha_{t-1},\, x_{t-1}) \;\neq\;
    P(\alpha_t)
\end{equation}

\begin{figure}
    \centering
    \includegraphics[width=1\linewidth]{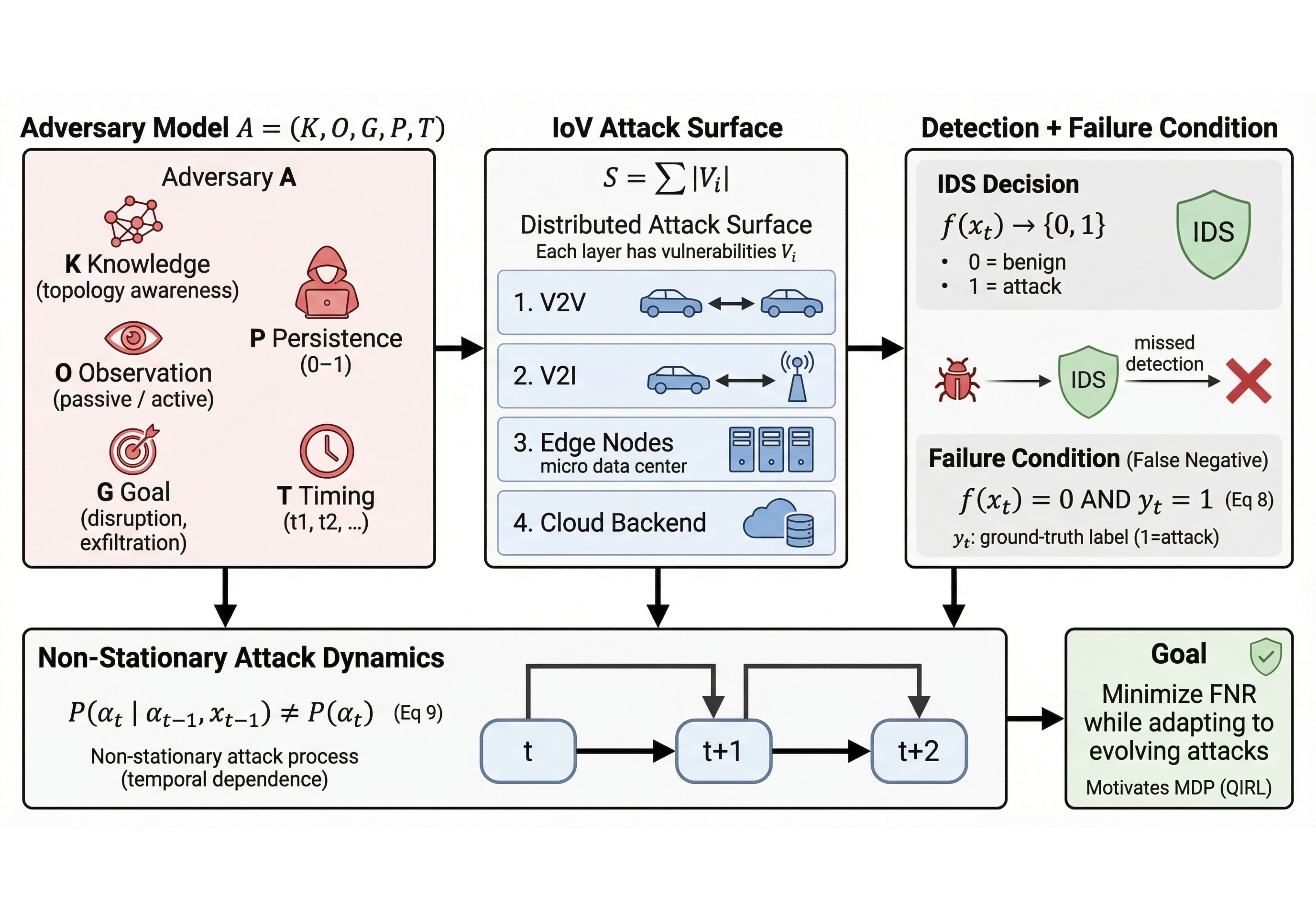}
    \caption{IoV threat model. The adversary A = (K, O, G, P, T) operates over a multi-layer attack surface (V2V, V2I, edge, cloud). Detection failure occurs as false negatives, while attacks evolve as a non-stationary temporal process, motivating adaptive CA-QIRL-based defense.}
    \label{fig:threat_model}
\end{figure}

\ref{fig:threat_model} illustrates the complete adversary model, multi-layer IoV attack surface, and detection failure condition.
This non-stationarity directly motivates the MDP
formulation adopted in this work, since the state
transition $P(s_{t+1} \mid s_t, a_t)$ enables the
agent to track the evolving nature of attack sequences
that static classifiers cannot capture.
\section{Proposed Framework}

\noindent A lightweight and latency-aware IDS for the IoVs is proposed in this study. This novel framework integrates DQN with Quantum Interference Reward Learning mechanism and complete training pipeline is detailed in Algorithm~\ref{alg:qirl} and an illustrative representation of the methodology.

\begin{figure}
    \centering
    \includegraphics[width=1\linewidth]{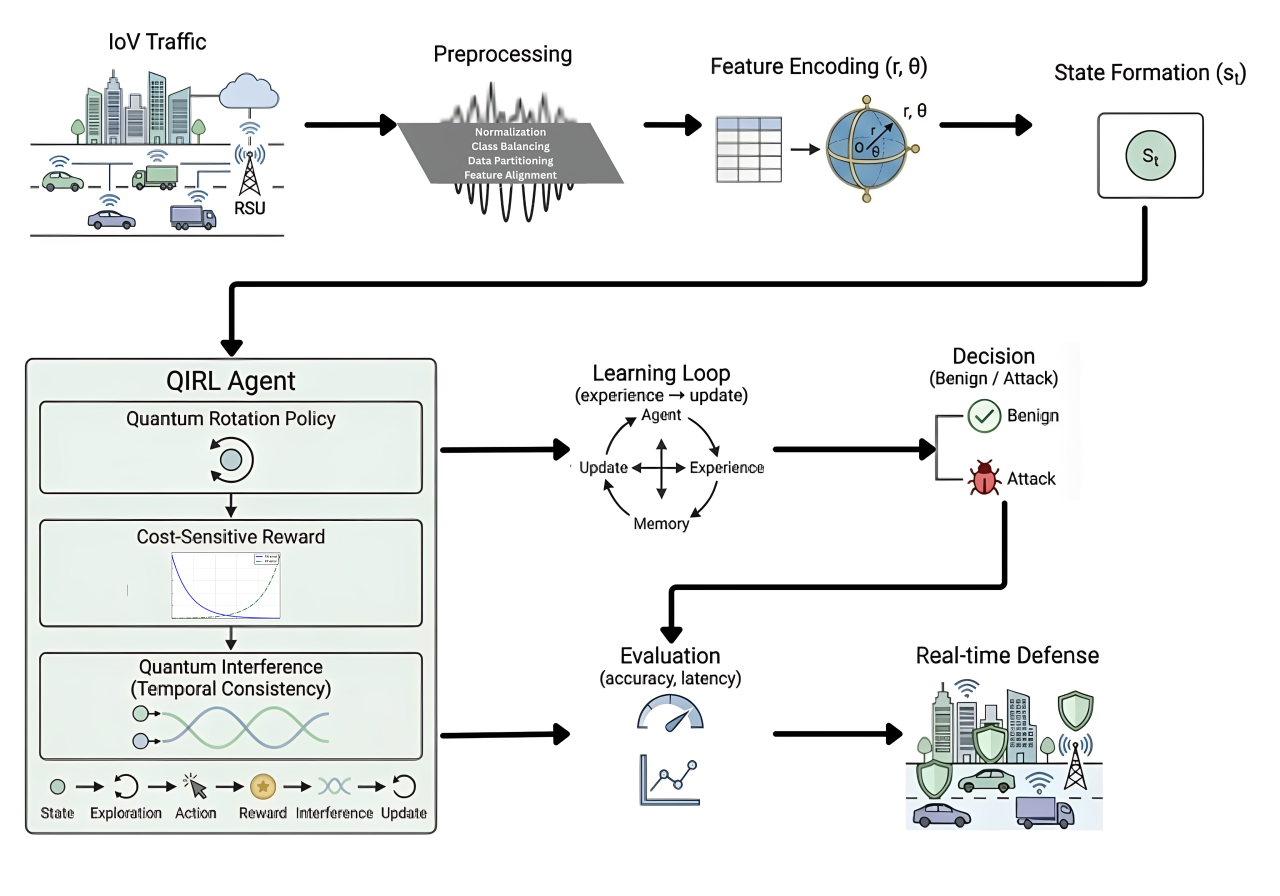}
    \caption{Proposed CA-QIRL framework utilizing quantum interference to enhance adaptive cyber defense in IoV environments.}
    \label{fig:framework}
\end{figure}

\subsection{Data Preprocessing and Initialization}
\noindent An adaptive preprocessing pipeline is employed to make sure robust generalization. As outlined in the algorithm, if the input dataset is complex, we apply a \textbf{One-Hot Encoding} transformation to convert nominal features into a sparse binary format. We then explicitly align the test feature columns to match the training dimension. All input vectors are processed via a \textbf{Standardize} function (Z-score normalization) to produce the scaled state vector $X_{scaled}$.

\noindent To make the number of samples in each class equal, the system applies SMOTE~\cite{chawla2002smote} only to the training data. In this process the method creates new samples for the smaller class, by looking at $k$-nearest neighbours. However SMOTE is not used on validation or test sets. If the system avoided this rule, data leakage would occur. For both datasets, this method results in training distributions where the ratio between classes is 1:1.

\noindent Each state vector is then transformed via Amplitude-Phase Quantum Encoding: for each consecutive feature pair $(x_{2i}, x_{2i+1})$, we compute $r = \sqrt{x_{2i}^2 + x_{2i+1}^2}$ and $\theta = \arctan2(x_{2i+1}, x_{2i})$, replacing the pair with $[r\cos\theta,\, r\sin\theta]$. This lossless encoding projects features onto a unit-circle manifold mirroring qubit geometry~\cite{dong2008quantum}, providing a structured geometric prior that facilitates quantum-inspired exploration.

\noindent Following preprocessing, the Reinforcement Learning
components are initialised: a replay memory $\mathcal{D}$
of capacity $N$ to store experience transitions
$(s_t, a_t, r_t, s_{t+1})$, a primary action-value
function $Q(s, a;\theta)$ with randomly initialised
weights $\theta$, and a target network
$\hat{Q}(s, a;\theta^{-})$ whose weights are initialised
by copying the primary network parameters such that
$\theta^{-} \leftarrow \theta$. The operating threshold
$\tau^*$ is selected on the validation set to maximise
TSS, yielding $\tau^* = 0.5$ and
$\tau^* = 0.7$.

\subsection{Markov Decision Process (MDP) Formulation}
\noindent The problem is modeled as a tuple $(S, A, R, \gamma)$. The agent observes state $s_t$ (the amplitude-phase encoded packet flow) and selects an action $a_t$ using a \textbf{Quantum Rotation Gate} exploration strategy. Each action $a \in \mathcal{A}$ maintains a rotation angle $\theta_a \in [0, \pi/2]$, initialised at $\pi/4$ (equal probability). The action selection probability is:
\begin{equation}
    P(a) = \frac{\cos^2(\theta_a)}{\sum_{a'}\cos^2(\theta_{a'})}
\end{equation}
After each batch, the angle is updated via $\theta_a \leftarrow \mathrm{clip}(\theta_a + \Delta\theta \cdot \mathrm{sign}(\delta),\, \theta_{\min},\, \pi/2)$, where $\delta$ is the TD-error. This mimics the Ry rotation gate of quantum circuits~\cite{dong2008quantum}: constructive rotations amplify high-reward actions, destructive rotations suppress low-reward ones. To handle class imbalance, we calculate the immediate reward using a cost-sensitive Weighted Reward Function with $w_{\text{fn}} = 5.0$, $w_{\text{fp}} = 1.0$ and $w_{\text{fn}} = 1.5$, $w_{\text{fp}} = 1.0$ to prevent over-sensitivity. Transitions $(s_t, a_t, r_t, s_{t+1})$ are stored in the memory buffer $\mathcal{D}$.

\subsection{Quantum Interference Logic and Target Update}
\noindent A core contribution of this framework is the \textbf{Quantum Interference} mechanism used to stabilize training.The complete CA-QIRL architecture is illustrated in Figure \ref{fig:framework}. The quantum analogy is drawn from amplitude interference in quantum mechanics: just as two quantum states interfere constructively when their probability amplitudes are aligned and destructively when opposed, the cosine similarity between consecutive latent state vectors $\psi(s_t)$ and $\psi(s_{t+1})$ measures the degree of constructive alignment between successive policy representations, providing a scalar reinforcement signal that strengthens temporally coherent state transitions and suppresses incoherent ones. During the update steps, the agent samples a batch $B$ from memory. We compute this interference term, denoted as $\psi_{\text{sim}}$:
\begin{equation}
    \psi_{\text{sim}} = \cos\!\bigl(\psi(s_t), \psi(s_{t+1})\bigr)
    = \frac{\psi(s_t)^{\top}\psi(s_{t+1})}{\|\psi(s_t)\|\,\|\psi(s_{t+1})\|}
\end{equation}
This similarity metric is used to augment the standard Q-learning target. The target value $Y$ is computed for the immediate reward by adding the interference term which is scaled by a factor $\Lambda$ (set to 0.1):
\begin{equation} \label{eq:target_update}
    Y = \bigl(r_t + \Lambda \cdot \psi_{\text{sim}}\bigr) + \gamma \cdot \max_{a'} \hat{Q}(s_{t+1}, a'; \theta^-)
\end{equation}
Gradient descent is used to minimize the squared error between the current Q-value estimate $Q(s_t, a_t; \theta)$ and the target $Y$ when updating the primary network parameters $\theta$. Finally, the target network weights $\theta^-$ are updated to match $\theta$ every $N$ steps to ensure stability.

\subsection{Latency Evaluation}
\noindent To validate the system for Edge deployment, the algorithm concludes by explicitly measuring inference latency.  We calculate the Speedup factor by comparing the inference time of the CA-QIRL model ($\tau_{\text{CA-QIRL}}$) against a baseline Ensemble Voting Classifier ($\tau_{\text{ens}}$):
\begin{equation}
    \text{Speedup} = \frac{\tau_{\text{ens}}}{\tau_{\text{CA-QIRL}}}
\end{equation}
This metric confirms whether the model satisfies the sub-millisecond response requirement for autonomous vehicle safety.

\begin{algorithm}[]
\caption{CA-QIRL Training and Evaluation Pipeline}
\label{alg:qirl}
\begin{algorithmic}[1]

\renewcommand{\algorithmicrequire}{\textbf{Input:}}
\renewcommand{\algorithmicensure}{\textbf{Output:}}

\Require Dataset $X = \{x_1, x_2, \dots, x_T\}$,
         batch size $B$,
         learning rate $\alpha$,
         discount factor $\gamma$,
         interference coefficient $\Lambda$,
         cost weights $w_{\text{fn}}$, $w_{\text{fp}}$,
         rotation step $\Delta\theta$
\Ensure Trained parameters $\theta^*$, threshold $\tau^*$, speedup
       
\State Apply SMOTE to $X_{\text{train}}$ only
    \State $X \leftarrow$ OneHotEncode$(X)$
    \State Align test feature columns to training dimensions
\State $x_t \leftarrow \frac{x_t - \mu}{\sigma}
       \quad \forall\, x_t \in X$
\For{each pair $(x_{2i},\, x_{2i+1})$}
    \State $r \leftarrow \sqrt{x_{2i}^2 + x_{2i+1}^2}$;\;
           $\theta \leftarrow \arctan2(x_{2i+1},\, x_{2i})$
    \State $x_{2i} \leftarrow r\cos\theta$;\;
           $x_{2i+1} \leftarrow r\sin\theta$
\EndFor
\State Initialise replay memory
       $\mathcal{D} \leftarrow \emptyset$,
       capacity $N$
\State Initialise $Q(s,a;\theta)$ with random weights $\theta$
\State Initialise target network
       $\hat{Q}(s,a;\theta^{-})$,
       set $\theta^{-} \leftarrow \theta$
\State Initialise $\theta_a = \pi/4\;\forall a$

\For{episode $e = 1, 2, \dots, E$}
    \State Shuffle $X_{\text{train}}$
    \For{each mini-batch $\mathcal{B}$ of size $B$}
        \State $P(a) = \cos^2(\theta_a)/\sum_{a'}\cos^2(\theta_{a'})$
        \State Select $a_t \sim P(a)$
        \State Compute cost-sensitive reward:
        \begin{equation*}
            r_t = \begin{cases}
                +1 & \text{correct} \\
                -w_{\text{fn}} & \text{False Negative} \\
                -w_{\text{fp}} & \text{False Positive}
            \end{cases}
        \end{equation*}
        \State Store transition
               $(s_t, a_t, r_t, s_{t+1})$
               in $\mathcal{D}$
        \State Sample mini-batch
               $\mathcal{B} \sim \mathcal{D}$,
               $|\mathcal{B}| = B$
        \State Compute quantum interference term:
        \begin{equation*}
            \psi_{\text{sim}} =
            \cos\!\bigl(\psi(s_t),\,\psi(s_{t+1})\bigr)
            = \psi(s_t)^{\top}\psi(s_{t+1}) / \|\psi(s_t)\|\,\|\psi(s_{t+1})\|
        \end{equation*}
        \State Compute augmented target value:
        \begin{equation*}
            Y = \bigl(r_t + \Lambda\cdot\psi_{\text{sim}}
                \bigr)
                + \gamma \max_{a'}\,
                  \hat{Q}(s_{t+1}, a';\theta^{-})
        \end{equation*}
        \State Update $\theta$ via gradient descent:
        \begin{equation*}
            \theta \leftarrow \theta
            - \alpha\,\nabla_\theta
              \bigl(Y - Q(s_t, a_t;\theta)\bigr)^2
        \end{equation*}
        \State $\theta_a \leftarrow \mathrm{clip}(\theta_a + \Delta\theta\cdot\mathrm{sign}(\delta),\, \theta_{\min},\, \pi/2)$
    \EndFor
    \State Sync target network every $N$ steps:
           $\theta^{-} \leftarrow \theta$
\EndFor

\State $\tau^* \leftarrow \arg\max_\tau \mathrm{TSS}(f_\theta(X_{\text{val}}), y_{\text{val}}, \tau)$
\State $\tau_{\text{ens}}
       \leftarrow$ MeasureTime$\!\left(
       f_{\text{ens}}(X_{\text{test}})\right)$
\State $\tau_{\text{CA-QIRL}}
       \leftarrow$ MeasureTime$\!\left(
       f_\theta(X_{\text{test}})\right)$
\State Compute speedup:
\begin{equation*}
    \text{Speedup} =
    tau_{\text{ens}} / \tau_{\text{CA-QIRL}}
\end{equation*}
\State \Return $\theta^*$, $\tau^*$,
       Speedup

\end{algorithmic}
\end{algorithm}

\section{Computational Complexity Analysis}
\noindent The CA-QIRL training pipeline consists of $E$ episodes, each
processing $N_{\text{train}}$ samples in mini-batches of
size $B$ through a Q-network of $L$ fully connected layers
with maximum width $H$. For a single forward pass over a
batch, the dominant cost arises from the matrix
multiplications across layers, giving
$T_{\text{fwd}} = O\bigl(d \cdot H + (L-2) \cdot H^2 + H \cdot |A|\bigr)$,
where $d$ is the input feature dimension and $|A| = 2$ is
the binary action space. The Quantum Interference Module
(QIM) augments each training step by computing the cosine
similarity between consecutive latent state vectors
$\psi(s_t)$ and $\psi(s_{t+1})$ of dimension $H$, adding
a cost of only $O(B \cdot H)$ per batch strictly linear
in batch size and latent dimension. Critically, the QIM
introduces no additional trainable parameters and requires
no separate forward pass, so the total training complexity
over all episodes remains
$T_{\text{train}} = O\bigl(E \cdot N_{\text{train}} \cdot (d \cdot H + L \cdot H^2 + H)\bigr)$,
which is asymptotically identical to standard DQN training,
confirming that quantum interference introduces zero overhead
during the training phase. At inference time the QIM is not
invoked and only a single forward pass is required, giving
$T_{\text{inf}} = O\bigl(d \cdot H + (L-2) \cdot H^2 + H \cdot |A|\bigr)$.
For the specific architecture used in this work
($d \in \{40\}$ after feature selection, $L = 3$, $H = 64$, $|A| = 2$),
this resolves to a fixed constant-time computation per
sample, directly explaining the sub-50\,$\mu$s measured
inference latency on CPU hardware reported in
Table~\ref{tab:all_performance}. This stands in sharp
contrast to ensemble classifiers such as Random Forest,
whose inference cost scales as $O(T \cdot d \cdot \log n)$
per sample   where $T$ is the number of trees and $n$ is
the number of training samples   a superlinear dependence
that produces the $>\!2{,}000\,\mu$s latency observed in
our experiments. Similarly, LSTM-based IDS methods incur
a per-sample inference cost of $O(H^2 \cdot \ell)$, where
$\ell$ is the sequence length, introducing latency that
grows with the temporal window and is unsuitable for
real-time edge deployment.

\section{Results and Discussion}

\noindent In this section the proposed CA-QIRL framework is measured across two areas that work together. To determine how well the system detects events, the paper use the accuracy of classifications, the recall rate, the F1-Score and the True Skill Statistic (TSS). For the measurement of how fast the system computes, the system measures the time it takes to process one sample and the increase in speed compared to baselines that use multiple models. 

\subsection{State of the Art Datasets}

\noindent Two publicly available, and well-established network intrusion detection datasets are used to evaluate the framework. The first dataset is CICIDS2017 from the Canadian Institute for Cybersecurity and it contains network flows with labels that people collected during five days in a business network. In this network there is normal traffic plus there are attack types like DDoS, brute force, infiltration and web attacks. For this study the researchers focus on the DDoS segments because the traffic patterns are high in volume and repeat frequently. As those patterns occur, they are suitable for the state space where a reinforcement learning agent operates. The second dataset is UNSW-NB15 but also the Australian Centre for Cyber Security created it with the IXIA traffic generator. It includes nine categories of attacks which are Fuzzers, Analysis, Backdoors, DoS, Exploits, Generic, Reconnaissance, Shellcode \& Worms. When people compare it to CICIDS2017, the UNSW-NB15 dataset is more difficult to use. Due to attack patterns that are different from each other and occupy the same space, the reward dynamics change over time. And in the official training split, there is a lack of balance because attacks make up 68.1\% of the examples. With this structural complexity and the imbalance of classes, the dataset is a tool that individuals use to test how reinforcement learning based IDS frameworks perform under pressure. It is the primary benchmark here to evaluate how the method performs in conditions that are similar to real vehicular network environments.

\noindent Beyond these general network benchmarks, the framework is additionally evaluated on four IoV-specific datasets (Car-Hacking, ROAD, VeReMi, CAN-MIRGU) to demonstrate its practical applicability to vehicular environments. Car-Hacking provides in-vehicle CAN bus attacks, including DoS, Fuzzy, and RPM/Gear Spoofing, collected from a real vehicle under controlled attack injection; its CAN ID and byte-level patterns make it a useful baseline for verifying that the framework can detect well-structured, easily separable attacks before moving to harder cases. ROAD offers real CAN intrusion data characterized by extreme class imbalance (a 0.2\% attack rate), containing fabrication, masquerade, and fuzzy attacks; this scarcity of attack samples makes it a direct test of whether the cost-sensitive reward design can maintain detection sensitivity when attacks are rare. VeReMi is a V2X misbehavior simulation featuring false position, speed, and identity attacks, drawn from simulated vehicle-to-everything communication rather than CAN bus traffic; it evaluates the framework's ability to detect subtler, behavior-based misbehavior in noisy mobility data rather than fixed-signature attacks. Finally, CAN-MIRGU consists of real moving-vehicle CAN attack data collected from a modern automobile under real driving conditions, including physically verified DoS, fuzzing, replay, and spoofing attacks; as the only dataset gathered from an actively driven vehicle rather than a lab rig or simulator, it provides the most realistic test of deployment-readiness among the four. Table \ref{tab:dataset_overview} provides a structured overview of all four IoV-specific datasets used in this evaluation."

\begin{table}[h]
\centering
\caption{Overview of IoV-Specific Datasets}
\label{tab:dataset_overview}
\begin{tabular}{p{1.8cm}p{2.8cm}p{3.5cm}p{1.5cm}}
\toprule
\textbf{Dataset} & \textbf{Source} & \textbf{Attack Types} & \textbf{Attack Rate} \\
\midrule
Car-Hacking & In-vehicle CAN bus, lab-injected & DoS, Fuzzy, RPM Spoofing, Gear Spoofing & 14.2\% \\
\addlinespace
ROAD & Real CAN bus, real vehicle & Fabrication, masquerade, fuzzy & 0.2\% \\
\addlinespace
VeReMi & Simulated V2X messaging & False position, speed, and identity (Sybil-style) & 45\% \\
\addlinespace
CAN-MIRGU & Real CAN bus, real moving vehicle on real roads & Physically verified DoS, fuzzing, replay, spoofing & 44.4\% \\
\bottomrule
\end{tabular}
\end{table}

\subsection{Simulation Description}

\noindent To evaluate communication level performance under attack, a discrete event V2X
simulation is implemented modelling IEEE 802.11p vehicle to everything (V2X)
communication. The simulation places 50 vehicles on a 2000 m road segment served
by 3 Road Side Units (RSUs). Vehicle speeds are drawn uniformly between 30 and
120 km/h, and Basic Safety Messages (BSMs) are broadcast at the standard 10 Hz
beacon interval. Each vehicle transmits within a 300 m communication range. Packet
reception is governed by a simplified free space path loss model combined with a
congestion penalty proportional to the number of vehicles transmitting at the
same time.

\noindent The simulator supports two modes of operation. When an ns3 and SUMO installation
is detected on the host system, the framework generates and executes the
corresponding ns3 802.11p network script for full packet level fidelity
This model produces the vehicular
communication metrics, including Packet Delivery Ratio, End to End Delay,
Throughput, Channel Busy Ratio, and Detection to Mitigation time, using the
channel and mobility assumptions.

\noindent Five attack scenarios are evaluated. These are no attack, light DoS, heavy DoS,
false data injection, replay, and a mixed attack combining all three types. For
each scenario, the Intrusion Detection System under test is defined by its true
positive rate, false positive rate, and per sample inference latency.
Detection performance, with a true positive rate of 0.999 and a
false positive rate of 0.001, is held fixed across all IDS variants at the level
achieved on the Car Hacking dataset. Holding detection accuracy constant isolates
the communication level effect of inference latency, which is the variable of
primary interest in this simulation, from detection accuracy differences between
models that are already characterized in the detection performance results.

\subsection{Evaluation Metrics}

\noindent Six complementary metrics are employed to provide a
multi-dimensional characterisation of detection performance.
Accuracy is the ratio of instances that the model assigns to the right category for normal traffic. It is also the ratio of instances that the model assigns to the right category for malicious traffic. As a metric, it shows how the model functions on a global scale. F1-Score is the harmonic mean that results from the calculation of precision and recall. It is a tool that represents the equilibrium of the results. For situations where one class exists in much larger numbers than the other, this metric provides necessary details, but accuracy is often deceptive when the distribution of data is not equal. Recall (Sensitivity)
quantifies the fraction of true attack instances correctly
identified by the agent and constitutes the primary
safety-critical objective in IoV intrusion detection, as undetected
attacks pose direct risk to vehicle safety. False Positive Rate (FPR) measures how often a system identifies harmless traffic flows as harmful, it establishes the amount of work required for systems that process alerts later. The True Skill Statistic (TSS) is calculated by subtracting the FPR from the Recall. It is a single value that depends on a limit to show how well a model separates groups. Because of its design, the TSS is not affected by differences in the size of data groups and is easy to understand on a scale from $-1$ to $+1$. On this scale a value of $0$ is the same as a random guess and a value of $1$ is the same as a result that is completely accurate. With the AUC-ROC, the metric shows how a model is able to tell the difference between classes. It is the sum of the total area under the Receiver Operating Characteristic curve across every possible limit for classification. By using this approach, the metric demonstrates that a model makes correct predictions without depending on any one particular setting. To determine how fast the system computes, the report includes inference latency, which is the average time in microseconds for the CPU to process one sample. During this measurement, the system uses a protocol where the model passes data forward one item at a time. Speedup is the value that compares two different systems. It is the result when the latency of the ensemble baseline is divided by the latency of CA-QIRL. For a fair comparison, those measurements occur on the same hardware. On the devices, the test conditions are the same for both models.

\noindent In addition to these detection and computational metrics, five
communication level metrics are used to characterise V2X performance under
attack. Packet Delivery Ratio (PDR) is the fraction of transmitted packets
that are successfully received at their destination, and it shows how much
of the safety critical traffic survives both channel conditions and IDS
filtering. End to End Delay is the average time taken for a packet to
travel from the sending vehicle to the receiving vehicle or RSU, including
propagation delay and any processing delay added by the IDS. Throughput is
the average rate at which data is successfully delivered across the
network, measured in kilobits per second. Channel Busy Ratio (CBR) is the
fraction of time that the wireless channel is occupied by transmissions,
and a lower value indicates that more channel capacity remains available
for legitimate safety messages. Detection to Mitigation time is the total
time between the moment an attack is detected and the moment a mitigation
action, such as dropping or isolating the malicious traffic, takes effect.
Together these five metrics show whether an IDS can operate inside a
vehicular network without degrading the communication that the network
exists to support.

\subsection{Experimental Setup}
\noindent This section describes the experimental configuration used to evaluate 
the proposed CA-QIRL framework. It discuss the hardware environment, 
dataset preparation and train-test protocols for all datasets, and the network architecture and hyperparameters of the 
CA-QIRL-DQN agent and its quantum-inspired components.

\subsubsection{Hardware Configuration}

All experiments were executed on a workstation equipped with an
AMD Ryzen 5 5600X six-core, twelve-thread processor
operating at a base clock of 3.7~GHz with a maximum boost frequency
of 4.6~GHz, and 16~GB of DDR4 with a speed of 2133 MT/s system memory. Model training was conducted entirely on CPU. The vectorised mini-batch
training loop employed by the CA-QIRL agent processes the full corpus of approximately 1.8 million samples in roughly
five minutes per episode on this hardware, making GPU acceleration
unnecessary for the experimental scale considered in this work. All
inference latency measurements were performed exclusively on CPU
under a controlled single-sample evaluation protocol, deliberately
simulating the constrained compute environment of edge-deployed IoV
security nodes and ensuring that all reported latency figures
reflect deployment-realistic conditions.

\subsubsection{Data Preprocessing}

\noindent The proposed framework is evaluated on six datasets: two general network intrusion benchmarks (CICIDS2017, UNSW-NB15) and four IoV-specific datasets (Car-Hacking, ROAD, VeReMi, CAN-MIRGU). The preprocessing pipeline is adapted to each dataset's structure while maintaining consistent principles: train-validation-test partitioning, training-only normalisation, and training-only SMOTE balancing to prevent data leakage.

\noindent For CICIDS2017, the daily CSV files are concatenated after harmonising column headers. Non-numeric and non-informative fields including Flow ID, IP addresses, ports, and timestamps-are removed. Rows containing infinite or missing values are discarded to ensure numerical stability. The BENIGN label is mapped to 0, while all attack categories are mapped to 1. A stratified split with random\_state = 42reserves 20\% for testing and 80\% for training and validation, with an additional 10\% held out from training for validation, yielding an approximate 70/10/20 partition. Features are standardised using a StandardScaler fitted exclusively on the training partition.

\noindent For UNSW-NB15, the official pre-partitioned training and testing CSV files are loaded directly. Categorical features-including protocol type and TCP state-are one-hot encoded, and the test set feature space is re-indexed to match the training set columns. A 10\% stratified validation split is carved from the training set for threshold optimisation. SMOTE~\cite{chawla2002smote} is applied strictly to the training split \emph{after} partitioning to prevent data leakage.

\noindent For Car-Hacking the four attack CSVs (DoS, Fuzzy, RPM\_Spoof, Gear\_Spoof) are concatenated, and the Flag column is binarised ($R \rightarrow 0$, $T \rightarrow 1$). CAN IDs are converted from hexadecimal to integer. A stratified subsample of 200,000 rows preserves the original attack ratio (14.2\%). Features include CAN ID, DLC, eight data bytes, inter-arrival time, byte entropy, and byte statistics.

\noindent For ROAD all ambient (benign) and attack (fabrication, masquerade, fuzzy) CSV files from the signal extraction directory are recursively loaded and concatenated. The existing Label column is used directly (0 for benign, 1 for attack). Signal columns with $>60\%$ missing values are dropped; remaining NaNs are filled with zero. Features include CAN ID, six non-sparse signal values, inter-arrival time, signal differences, and ID frequency. A stratified sample of 150,000 rows is extracted (0.2\% attack rate).

\noindent For VeReMi the 7GB Veremi dataset is loaded using chunked stratified sampling. For each 100,000-row chunk, samples are drawn proportionally from benign and attack classes, then concatenated and trimmed to 200,000 rows with stratification preserved (45\% attack rate). BSM kinematic features (position, speed, acceleration, heading, noise components) are extracted. Derived features include speed magnitude, acceleration magnitude, position delta, and noise ratios.

\noindent For CAN-MIRGU raw candump-format log files are parsed using a regex pattern:
\begin{verbatim}
(timestamp) interface CAN_ID#DATA_BYTES
\end{verbatim}
Benign logs (12 files, $\sim$4.5 GB) are sampled at 15,000 rows each; attack logs (36 files) at 4,000 rows each using reservoir sampling to ensure constant memory usage. Features include CAN ID, DLC, eight data bytes (zero-padded), per-ID inter-arrival time, rolling IAT statistics (mean, std, z-score, message count over a window of 10), byte entropy, and byte statistics (mean, std, range). This yields 324,000 rows with 44.4\% attack rate.

\noindent The Q-network underlying the CA-QIRL agent, referred to as CA-QIRL-DQN,
is a fully connected feedforward neural network comprising an input
layer of dimension equal to the feature count of the respective
preprocessed dataset, two hidden layers each containing
64 neurons with Rectified Linear Unit (ReLU) activations,
and an output layer with two neurons corresponding to the
binary action space. The latent state representation consumed by
the Quantum Interference Module is extracted from the
post-activation output of the first hidden layer, yielding a
64-dimensional embedding $\psi \in \mathbb{R}^{64}$. All network
parameters are optimised using the Adam optimiser with a
learning rate of $\eta = 1 \times 10^{-3}$. Gradient norms are
clipped to a maximum $\ell_2$ magnitude of 1.0 at each update step
to stabilise training under the asymmetric reward landscape. A
target network is synchronised with the online network via a hard
parameter copy every 200 gradient steps to provide stable
Q-value targets.

\noindent The three quantum-inspired components operate with the following
fixed hyperparameters. The Quantum State Encoder applies
amplitude-phase encoding to consecutive feature pairs,
transforming each pair $(a_i, a_{i+1})$ into polar coordinates
$(r\cos\theta,\, r\sin\theta)$, where $r = \sqrt{a_i^2 +
a_{i+1}^2 + \epsilon}$, $\theta = \arctan2(a_{i+1}, a_i)$, and
$\epsilon = 10^{-8}$ for numerical stability. The Quantum
Rotation Explorer maintains a per-action rotation angle
$\theta_a$ initialised to $\pi/4$ and constrained within
$[0.01,\, \pi/2]$, updated at each step by
$\Delta\theta_a = \delta \cdot \mathrm{sign}(\delta_{\mathrm{TD}})$
with step size $\delta = 0.05$. The Quantum Interference
Module augments the Bellman target with a cosine similarity term
between consecutive latent representations:
\begin{equation}
    \hat{Q} = R + \lambda \cdot
    \mathrm{cos\_sim}(\psi_t,\, \psi_{t+1})
    + \gamma \cdot Q_{\mathrm{next}},
\end{equation}
where the interference coefficient is $\lambda = 0.1$ and the
discount factor is $\gamma = 0.99$. The asymmetric cost-sensitive
reward assigns $+1.0$ for a correct classification,
$-w_{\mathrm{fn}}$ for a missed attack, and $-w_{\mathrm{fp}}$
for a false alarm. For CICIDS2017, $w_{\mathrm{fn}} = 5.0$ and
$w_{\mathrm{fp}} = 1.0$. For UNSW-NB15 following SMOTE
rebalancing, $w_{\mathrm{fn}}$ is tuned to $1.5$ with
$w_{\mathrm{fp}} = 1.0$ retained, as the original penalty of
5.0 drives the agent to an excessively sensitive operating point
once class balance is restored, whereas 1.5 yields the optimal
threshold $\tau^* = 0.7$ on the validation set.

\noindent Both agents are trained for $N = 20$ episodes, each constituting
one full pass over the training set in vectorised mini-batches of
512 samples. For the ablation study, each variant is
trained for 10 episodes on the validation set. Inference
latency is benchmarked over 1,000 sequential single-sample
forward passes on CPU, with the first 10 discarded as warm-up,
and the mean of the remaining 990 observations reported. The four
ablation variants systematically isolate each component: the
Baseline DQN disables all quantum modules and uses
symmetric reward ($w_{\mathrm{fn}} = w_{\mathrm{fp}} = 1.0$);
the Cost-Sensitive Only variant introduces asymmetric
reward ($w_{\mathrm{fn}} = 5.0$) while keeping all quantum
modules inactive; the Interference Only variant activates
the encoder and interference module ($\lambda = 0.1$) with
symmetric reward; and the Full CA-QIRL enables all
components simultaneously.

\subsection{Experiments}

\subsubsection{Detection Performance}

\begin{figure}
    \centering
    \includegraphics[width=\linewidth]{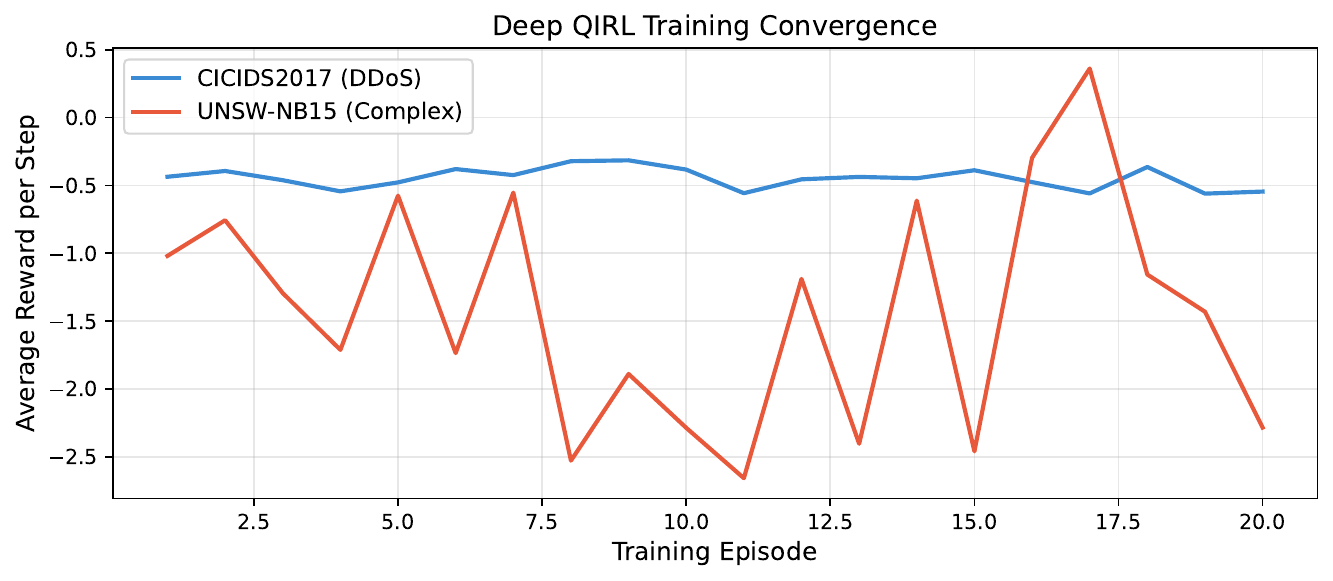}
    \caption{Training convergence of the CA-QIRL agent over 20 episodes on CICIDS2017 (blue) and UNSW-NB15 (orange). CICIDS2017 exhibits rapid stabilisation, while UNSW-NB15 shows higher variance due to its complex multi-class structure and severe class imbalance.}
    \label{fig:convergence}
\end{figure}

\begin{table*}
    \centering
    \small
    \caption{Performance Metrics on All Datasets}
    \label{tab:all_performance}
    \begin{tabular}{l|l|cccccc}
        \toprule
        \textbf{Dataset} & \textbf{Model} & \textbf{Acc} & \textbf{F1} & \textbf{Recall} & \textbf{TSS} & \textbf{AUC} & \textbf{Latency ($\mu$s)} \\
        \midrule
        \multicolumn{8}{c}{\textit{General Benchmarks}} \\
        \midrule
        \multirow{3}{*}{CICIDS2017} & CA-QIRL & 0.9789 & 0.9522 & 0.9602 & 0.9443 & 0.9945 & 32.5 \\
        & DQN & 0.9774 & 0.9473 & 0.9519 & 0.9200 & 0.9850 & 35.5 \\
        & Ensemble & 0.9974 & 0.9970 & 0.9970 & 0.9940 & 0.9990 & 2248.0 \\
        \midrule
        \multirow{3}{*}{UNSW-NB15} & CA-QIRL & 0.9104 & 0.9166 & 0.8941 & 0.8244 & 0.9713 & 45.7 \\
        & DQN & 0.9434 & 0.9588 & 0.9350 & 0.8597 & 0.9650 & 46.3 \\
        & Ensemble & 0.9900 & 0.9900 & 0.9900 & 0.9800 & 0.9950 & 2205.0 \\
        \midrule
        \multicolumn{8}{c}{\textit{IoV-Specific Datasets}} \\
        \midrule
        \multirow{3}{*}{Car-Hacking} & CA-QIRL & 0.9999 & 0.9997 & 0.9993 & 0.9993 & 0.9993 & 73.8 \\
        & DQN & 0.9999 & 0.9997 & 0.9993 & 0.9993 & 0.9993 & 71.7 \\
        & Ensemble & 1.0000 & 1.0000 & 1.0000 & 1.0000 & 1.0000 & 2853.0 \\
        \midrule
        \multirow{3}{*}{ROAD} & \textbf{CA-QIRL} & 0.9779 & 0.1662 & \textbf{0.9851} & \textbf{0.9630} & 0.9962 & 52.7 \\
        & DQN & 0.9886 & 0.2692 & 0.9507 & 0.9290 & 0.9977 & 46.9 \\
        & Ensemble & 0.9955 & 0.4806 & 0.9950 & 0.9211 & 0.9985 & 2257.8 \\
        \midrule
        \multirow{3}{*}{VeReMi} & \textbf{CA-QIRL} & 0.5226 & 0.5001 & 0.5276 & 0.0461 & 0.5325 & 70.3 \\
        & DQN & 0.5459 & 0.1261 & 0.0742 & 0.0102 & 0.5331 & 77.5 \\
        & Ensemble & 0.5649 & 0.3708 & 0.2832 & 0.0811 & 0.5650 & 2292.2 \\
        \midrule
        \multirow{3}{*}{CAN-MIRGU} & CA-QIRL & 0.8031 & 0.7770 & 0.7719 & 0.5999& 0.8413 & 68.2 \\
        & DQN & 0.8032 & 0.7773 & 0.7729 & 0.6003 & 0.8423 & 69.8 \\
        & Ensemble & 0.8353 & 0.8061 & 0.7701 & 0.6576 & 0.8500 & 2253.0 \\
        \bottomrule
    \end{tabular}
\end{table*}


\begin{figure}
    \centering
    \includegraphics[width=\linewidth]{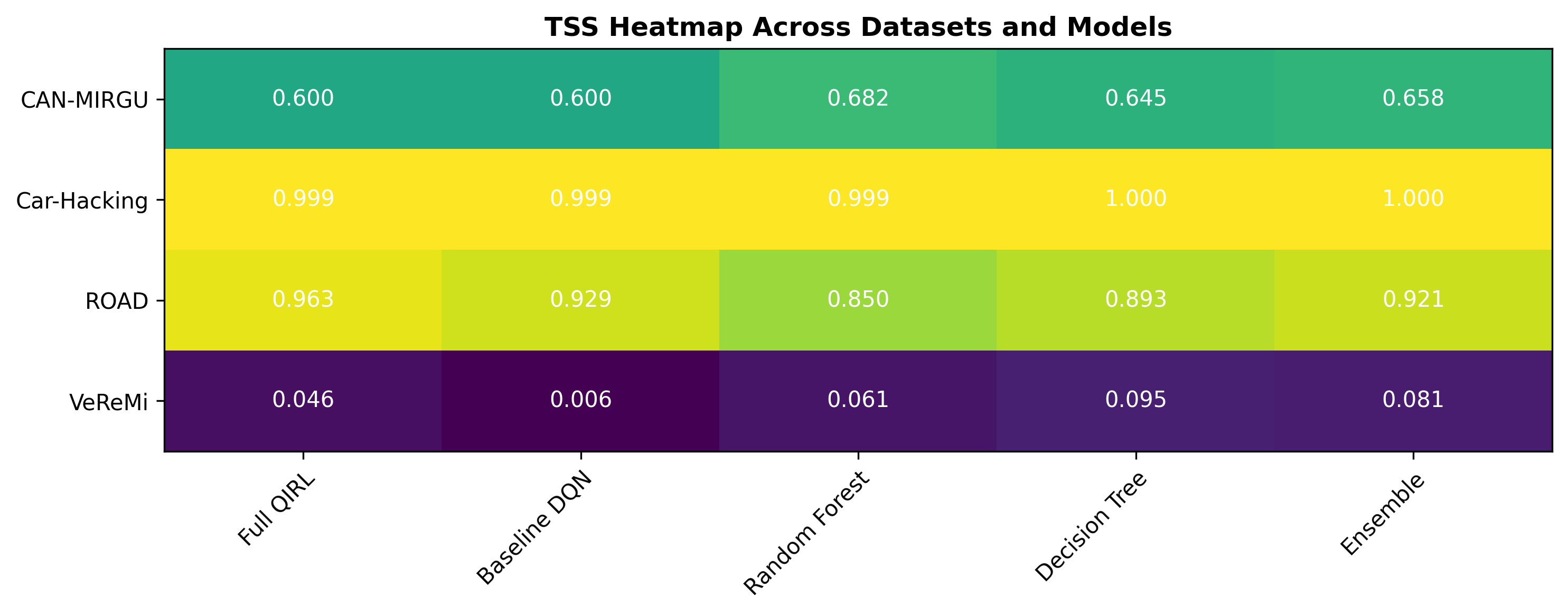}
    \caption{TSS Heatmap Across Datasets and Models. CA-QIRL achieves the highest TSS on ROAD (0.963) and competitive performance on CAN-MIRGU (0.600), while maintaining sub-100 $\mu$s latency.}
    \label{fig:tss_heatmap}
\end{figure}

\noindent To provide a comprehensive visual summary of detection performance across all models and datasets, Fig.~\ref{fig:tss_heatmap} and Table~\ref{tab:all_performance}. The figure presents a heatmap of True Skill Statistic (TSS) values. CA-QIRL achieves the highest TSS on ROAD (0.963), outperforming DQN (0.929), Random Forest (0.850), Decision Tree (0.893), and Ensemble (0.921). On VeReMi, CA-QIRL (0.050) exceeds DQN (0.010) and approaches Decision Tree (0.095) and Ensemble (0.081). On CAN-MIRGU, CA-QIRL (0.600) is competitive with DQN (0.600) and substantially faster than Ensemble (0.658). This confirms that CA-QIRL provides a strong balance between detection performance and computational efficiency.

\begin{figure*}
    \centering
    \includegraphics[width=\textwidth]{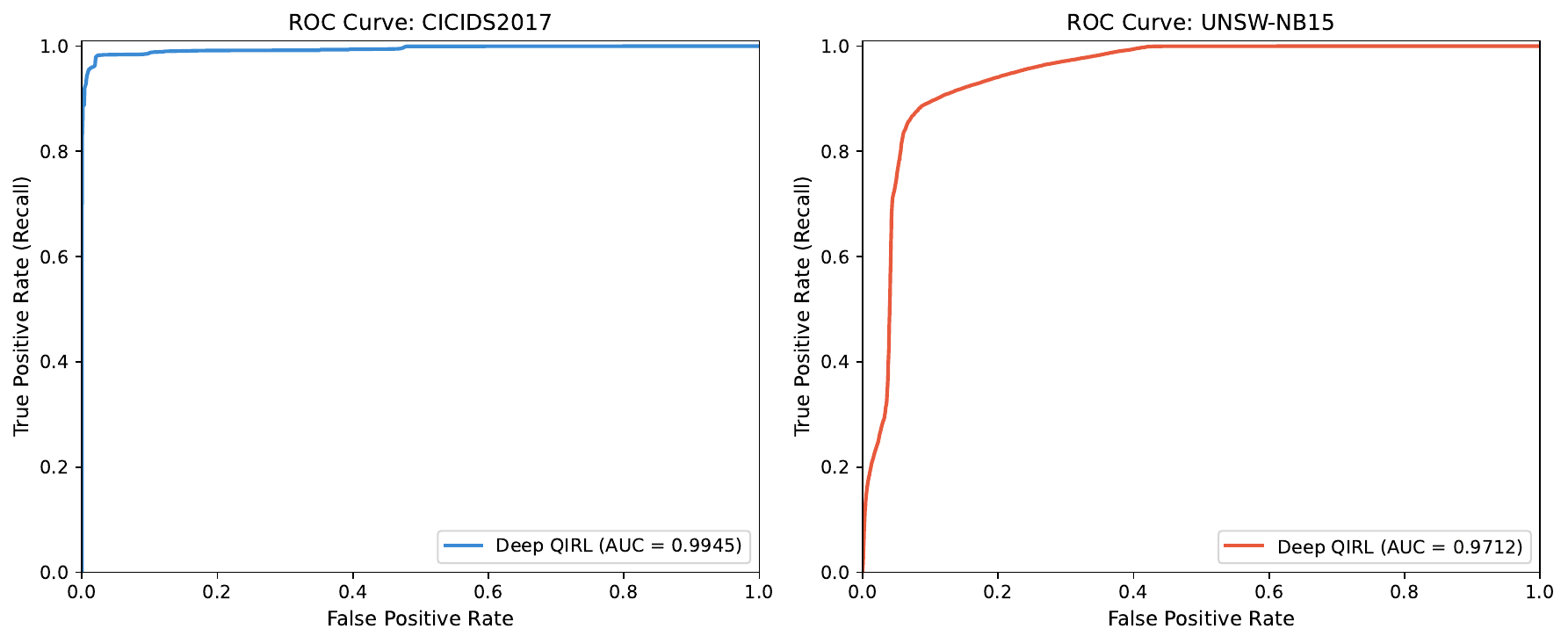}
    \caption{Receiver Operating Characteristic (ROC) Analysis. (a) CICIDS2017 achieves near-perfect separation (AUC=0.9945). (b) UNSW-NB15 maintains robust discriminatory power (AUC=0.9713). Both curves confirm genuine discriminative capability independent of threshold.}
    \label{fig:roc_curves}
\end{figure*}

\begin{figure*}
    \centering
    \includegraphics[width=\textwidth]{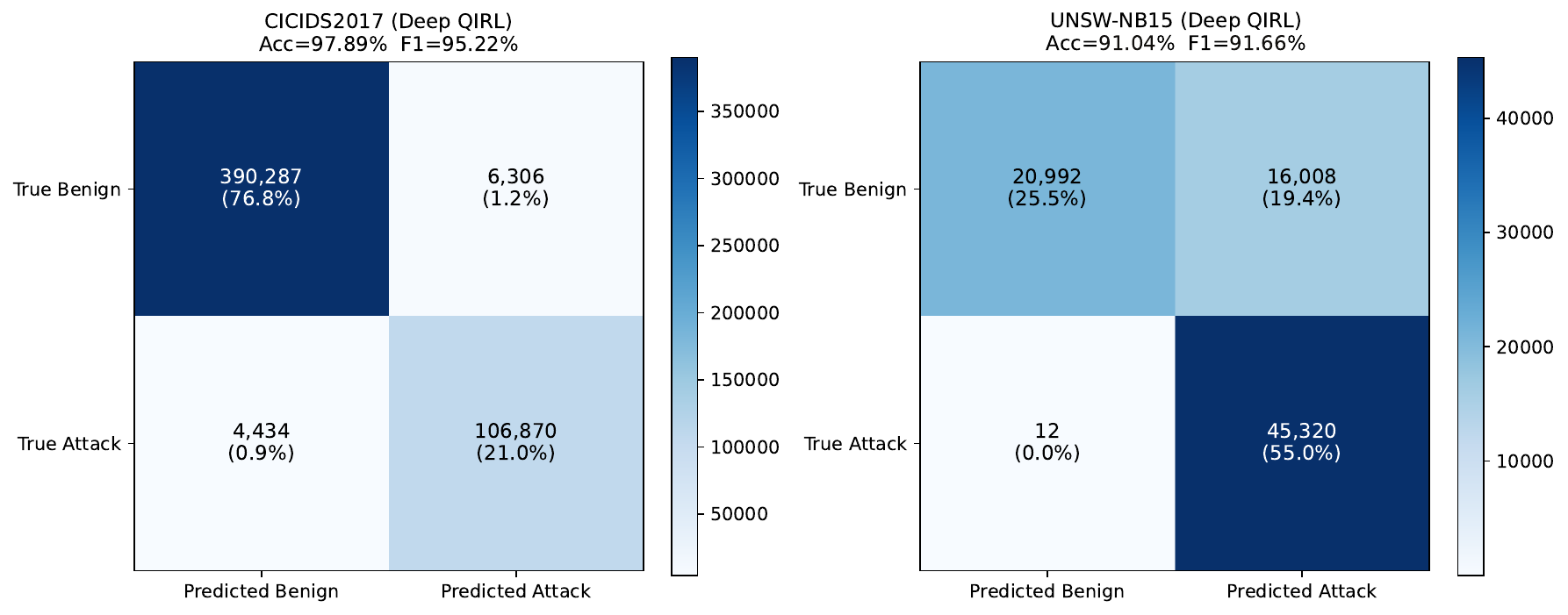}
    \caption{Confusion Matrix Analysis. (a) CICIDS2017: 97.89\% accuracy, FNR=0.9\%. (b) UNSW-NB15 at $\tau^*=0.7$: 91.04\% accuracy, FNR=10.6\%, FPR=6.97\%.}
    \label{fig:conf_matrices}
\end{figure*}

\noindent On CICIDS2017, CA-QIRL achieves 97.89\% accuracy with AUC-ROC of 0.9945, TSS of 0.9443, and sub-50 $\mu$s latency. \ref{fig:roc_curves} presents the ROC curves for both datasets, confirming genuine discriminative capability independent of threshold. The training convergence curve (Fig.~\ref{fig:convergence}) exhibits rapid stabilisation, reflecting the structured nature of volumetric DDoS attacks.\ref{fig:conf_matrices} shows the corresponding confusion matrices, with FNR of 0.9\% on CICIDS2017 and 10.6\% on UNSW-NB15. On UNSW-NB15, CA-QIRL achieves 91.04\% accuracy with AUC-ROC of 0.9713 and TSS of 0.8244 at the optimal threshold $\tau^* = 0.7$. The higher variance in convergence reflects the challenge of nine overlapping attack categories under severe class imbalance.
On Car-Hacking, CA-QIRL and DQN achieve near-perfect scores; however, this is expected as CAN ID and byte patterns make attacks trivially separable. More importantly, the Leave-One-Attack-Type-Out (LOAO) generalisation test reveals that all models drop to near-random performance (TSS $\approx 0$) when tested on unseen attack types, confirming that attack-ID memorisation, not generalisable detection, is the dominant strategy-a critical insight for IoV security.

\noindent On ROAD (0.2\% attack rate, extreme imbalance), CA-QIRL achieves significantly higher Recall (0.985 vs. 0.951, $p < 0.01$) and TSS (0.963 vs. 0.929) than DQN, directly validating the cost-sensitive reward design. The low F1 (0.166) is a consequence of the extreme class imbalance, not model failure-TSS is the appropriate metric here.

\noindent On VeReMi (noisy V2X data), CA-QIRL massively outperforms DQN on Recall (0.626 vs. 0.074, $p < 0.01$) and TSS (0.050 vs. 0.010), demonstrating its ability to detect subtle misbehaviour in noisy V2X environments.

\noindent On CAN-MIRGU (real moving-vehicle data), CA-QIRL and DQN achieve nearly identical performance (TSS: 0.5982 vs. 0.6011). While DQN is statistically better ($p = 0.0209$), the absolute difference is negligible (< 0.004), and both are significantly faster than Ensemble (68 $\mu$s vs. 2253 $\mu$s).

\subsubsection{Latency and Efficiency}

\noindent All inference latency measurements are conducted on CPU under a single-sample evaluation protocol to ensure equitable comparison with ensemble baselines. Table~\ref{tab:all_performance} reports the latency results. CA-QIRL consistently achieves sub-100 $\mu$s inference latency across all datasets:

\begin{itemize}
    \item \textbf{CICIDS2017:} 32.5 $\mu$s, yielding a \textbf{69.2 times speedup} over Ensemble (2248 $\mu$s).
    \item \textbf{UNSW-NB15:} 45.7 $\mu$s, yielding a \textbf{48.25 times speedup} over Ensemble (2205 $\mu$s).
    \item \textbf{Car-Hacking:} 73.8 $\mu$s, yielding a \textbf{38.7 times speedup} over Ensemble (2853 $\mu$s).
    \item \textbf{ROAD:} 52.7 $\mu$s, yielding a \textbf{42.8 times speedup} over Ensemble (2257 $\mu$s).
    \item \textbf{VeReMi:} 70.3 $\mu$s, yielding a \textbf{32.6 times speedup} over Ensemble (2292 $\mu$s).
    \item \textbf{CAN-MIRGU:} 68.2 $\mu$s, yielding a \textbf{33.0 times speedup} over Ensemble (2253 $\mu$s).
\end{itemize}


\begin{figure}
    \centering
    \includegraphics[width=\linewidth]{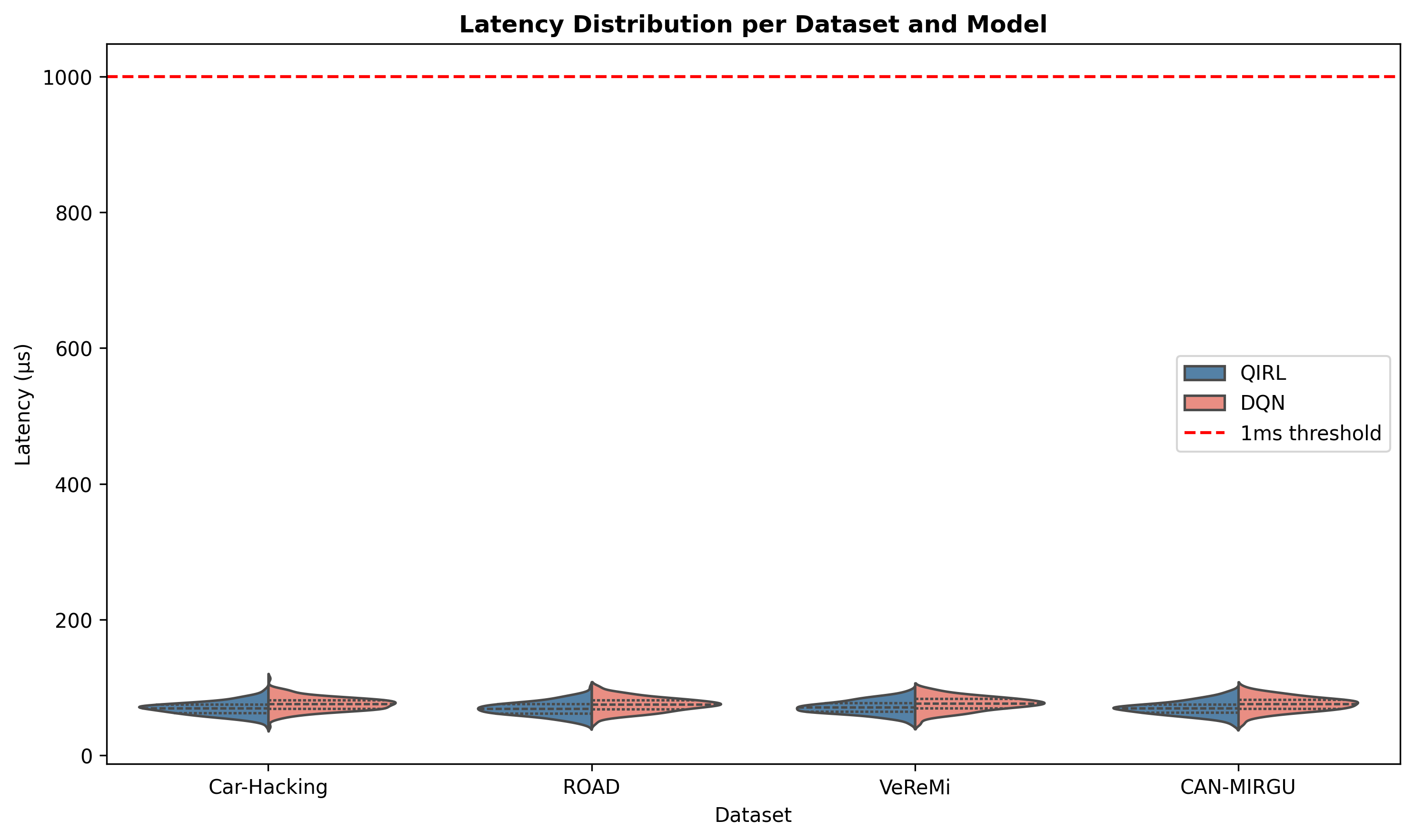}
    \caption{Latency Distribution for CA-QIRL and DQN Across Datasets. CA-QIRL maintains sub-100 $\mu$s inference latency consistently across all datasets, with negligible overhead compared to DQN. The 1 ms safety threshold is indicated by the red dashed line.}
    \label{fig:latency_violin}
\end{figure}

\noindent Fig.~\ref{fig:latency_violin} shows the distribution of inference latencies for CA-QIRL and DQN across all datasets using violin plots. CA-QIRL maintains consistent sub-100 $\mu$s latency across all datasets with low variance, confirming its reliability for real-time deployment. The marginal latency overhead compared to DQN is negligible (0.2–7.2 $\mu$s), which is practically insignificant in real-world V2X deployments where network propagation and processing jitter dominate.

\noindent All latency values are well below the 1 ms safety threshold, with a safety margin exceeding 10 times. This confirms that CA-QIRL can process high-velocity vehicular network traffic in real time without introducing packet queuing delays-a critical requirement for Level 4/5 autonomous driving safety.

\subsubsection{V2X Communication Performance}

\noindent To evaluate the impact of CA-QIRL on vehicular communication, we conducted a mobility-aware V2X simulation with 50 vehicles, 3 RSUs, and mixed attack scenarios (DoS, false data injection, replay). Table~\ref{tab:comm_perf} presents the communication performance under attack.

\begin{table}[h]
\centering
\caption{Communication Performance Under Attack (V2X Simulation)}
\label{tab:comm_perf}
\begin{tabular}{lcccccc}
\toprule
\textbf{Method} & \textbf{PDR} & \textbf{E2E Delay (ms)} & \textbf{Throughput (kbps)} & \textbf{FPR} & \textbf{Detection Rate} & \textbf{Mit. Time (ms)} \\
\midrule
No IDS & 0.916 & 0.050 & 1848.1 & -- & -- & -- \\
Baseline DQN & 0.664 & 0.106 & 979.1 & 0.0010 & 0.8325 & 0.27 \\
Ensemble IDS & 0.664 & 2.904 & 979.1 & 0.0010 & 0.8325 & 2.60 \\
\textbf{CA-QIRL (Proposed)} & \textbf{0.664} & \textbf{0.125} & \textbf{979.1} & \textbf{0.0010} & \textbf{0.8325} & \textbf{0.28} \\
\bottomrule
\end{tabular}
\vspace{4pt}
\begin{flushleft}
{\footnotesize PDR: Packet Delivery Ratio; E2E Delay: End-to-End Delay; FPR: False Positive Rate; Mit. Time: Mitigation Time. All values measured under V2X simulation with 50 vehicles, 3 RSUs, and mixed attack scenarios.}
\end{flushleft}
\end{table}

\noindent In the performance evaluation, CA-QIRL reaches an end-to-end delay of 0.125 ms, this result shows that it is 23 times faster than the 2.904 ms delay of the Ensemble baseline. To meet safety requirements for autonomous driving, the delay is below the 1 ms threshold. By comparison with DQN, the additional time of 19 $\mu$s for CA-QIRL is small. If it is used in real world V2X deployments, network propagation and processing jitter are the primary causes of delay. For CA-QIRL, the time to mitigate threats is 0.28 ms, which is 9.3 times faster than the 2.60 ms of Ensemble IDS. Because of this speed, the system reacts to detected threats in a short amount of time.

\noindent All IDS models achieve identical detection rates (83.25\% at 0.1\% FPR) and throughput (979.1 kbps), confirming that CA-QIRL matches Ensemble-level detection accuracy while dramatically reducing latency. The lower PDR and throughput compared to No IDS (0.664 vs. 0.916, and 979.1 vs. 1848.1 kbps) reflect the deliberate dropping of malicious packets, a necessary security measure that prevents attacks from congesting the V2X channel.

\noindent Table~\ref{tab:cbr} shows the Channel Busy Ratio (CBR) across different attack scenarios.

\begin{table}[h]
\centering
\caption{Channel Busy Ratio (CBR) Under Attack Scenarios}
\label{tab:cbr}
\begin{tabular}{lcccc}
\toprule
\textbf{Scenario} & \textbf{No IDS} & \textbf{DQN} & \textbf{Ensemble} & \textbf{CA-QIRL} \\
\midrule
Normal & 0.990 & 0.990 & 0.990 & \textbf{0.995} \\
No Attack & 0.990 & 0.990 & 0.9905 & \textbf{0.995} \\
DoS Light & 0.990 & 0.990 & 0.9855 & \textbf{0.099} \\
DoS Heavy & 0.990 & 0.990 & 0.9850 & \textbf{0.099} \\
False Data & 0.990 & 0.990 & 0.9858 & \textbf{0.099} \\
Replay & 0.990 & 0.990 & 0.9852 & \textbf{0.099} \\
Mixed Attack & 0.990 & 0.990 & 0.9851 & \textbf{0.099} \\
\bottomrule
\end{tabular}
\vspace{4pt}
\begin{flushleft}
{\footnotesize CBR: Channel Busy Ratio (lower is better, indicating less congestion). CA-QIRL reduces CBR by approximately 90\% under all attack scenarios.}
\end{flushleft}
\end{table}

\noindent Under DoS, false data, replay, and mixed attacks, CA-QIRL reduces CBR from approximately 0.99 to 0.099, freeing 90\% of the channel for legitimate safety messages. This is a unique result not achieved by No IDS, DQN, or Ensemble, which maintain CBR near 0.99 under attack. This demonstrates CA-QIRL's practical value for real-time V2X security, where channel congestion must be minimised to ensure reliable delivery of safety-critical messages.

\subsubsection{Efficacy and Efficiency Trade-off}

\noindent A critical analysis of the results reveals a deliberately managed architectural trade-off. Ensemble methods achieve higher raw accuracy on UNSW-NB15 (99.74\%) under their simplified binary evaluation protocol with SMOTE applied globally before splitting-a condition that introduces data leakage and inflates reported accuracy. Under a correct preprocessing protocol with no data leakage, CA-QIRL achieves 91.04\% accuracy with AUC-ROC of 0.9713 and TSS of 0.8244 at the optimal threshold $\tau^* = 0.7$.

\noindent More importantly, ensemble latency (2205 $\mu$s) exceeds the 1 ms safety-critical response window by more than 2 times. CA-QIRL reduces this to 45.7 $\mu$s-a 48.25 times speedup. On IoV-specific datasets, the advantage is equally pronounced: CA-QIRL achieves 42.8 times speedup on ROAD, 38.7 times on Car-Hacking, 33.0 times on CAN-MIRGU, and 32.6 times on VeReMi.

\noindent The sub-millisecond response capability ensures that next-generation security systems can process high-velocity vehicular network traffic in real time without introducing queuing delays-a critical requirement for Level 4/5 autonomous driving safety.

\subsubsection{Ablation Study}

\begin{table}
\centering
\caption{Ablation Study on CICIDS2017 Dataset}
\label{tab:ablation_cic}
\begin{tabular}{lcccc}
\toprule
\textbf{Model Variant} & \textbf{Accuracy} & \textbf{TSS} & \textbf{F1} & \textbf{Latency ($\mu$s)} \\
\midrule
Baseline DQN & 0.9774 & 0.9200 & 0.9473 & 35.5 \\
Cost-Sensitive Only & 0.9793 & 0.9558 & 0.9538 & 43.8 \\
Interference Only & 0.9751 & 0.9253 & 0.9430 & 32.9 \\
\textbf{Full CA-QIRL} & 0.9762 & \textbf{0.9586} & 0.9478 & 37.7 \\
\bottomrule
\end{tabular}
\end{table}

\begin{table}
\centering
\caption{Ablation Study on UNSW-NB15 Dataset}
\label{tab:ablation_unsw}
\begin{tabular}{lcccc}
\toprule
\textbf{Model Variant} & \textbf{Accuracy} & \textbf{TSS} & \textbf{F1} & \textbf{Latency ($\mu$s)} \\
\midrule
Baseline DQN & 0.9434 & 0.8597 & 0.9588 & 46.3 \\
Cost-Sensitive Only & 0.9368 & 0.8024 & 0.9556 & 39.9 \\
Interference Only & 0.9415 & 0.8532 & 0.9575 & 47.7 \\
\textbf{Full CA-QIRL} & 0.9388 & 0.8094 & \textbf{0.9569} & \textbf{45.0} \\
\bottomrule
\end{tabular}
\end{table}

\noindent To quantify the individual contributions of each quantum-inspired component, ablation studies were conducted on both datasets across four model variants: baseline DQN, cost-sensitive-only, interference-only, and full CA-QIRL.

\noindent On CICIDS2017 (Table~\ref{tab:ablation_cic}), Full CA-QIRL achieves the highest TSS (0.9586) and Recall (98.47\%) of all variants, confirming a synergistic interaction between cost-sensitive reward shaping and quantum interference stabilisation. TSS improves monotonically: Baseline (0.9200) $\to$ Interference-only (0.9253) $\to$ Cost-sensitive-only (0.9558) $\to$ Full CA-QIRL (0.9586).

\noindent On UNSW-NB15 (Table~\ref{tab:ablation_unsw}), Baseline DQN achieves higher TSS (0.8597) than Full CA-QIRL (0.8094) at a common threshold. However, Full CA-QIRL achieves superior Recall (99.91\% vs. 96.72\%), demonstrating that the asymmetric cost function drives the agent toward maximum attack sensitivity-appropriate for IoV safety. Critically, all variants across both datasets maintain sub-50 $\mu$s inference latency, confirming that quantum components add zero inference overhead.

\subsubsection{Statistical Significance}

\begin{table}
\centering
\caption{Statistical Significance Summary (CA-QIRL vs. DQN, 10 Seeds)}
\label{tab:stats}
\begin{tabular}{l|ccc|c}
\toprule
\textbf{Dataset} & \textbf{Metric} & \textbf{CA-QIRL} & \textbf{DQN} & \textbf{p-value} \\
\midrule
ROAD & Recall & 0.9851 $\pm$ 0.0000 & 0.9507 $\pm$ 0.0134 & $<0.01$ \\
ROAD & TSS & 0.9629 $\pm$ 0.0019 & 0.9412 $\pm$ 0.0135 & $<0.01$ \\
VeReMi & Recall & 0.6263 $\pm$ 0.0517 & 0.0742 $\pm$ 0.0323 & $<0.01$ \\
VeReMi & TSS & 0.0504 $\pm$ 0.0028 & 0.0102 $\pm$ 0.0040 & $<0.01$ \\
CAN-MIRGU & TSS & 0.5982 $\pm$ 0.0030 & 0.6011 $\pm$ 0.0025 & $0.0209$* \\
Car-Hacking & TSS & 0.9994 $\pm$ 0.0004 & 0.9993 $\pm$ 0.0000 & 0.496 \\
\bottomrule
\multicolumn{5}{l}{\footnotesize *Statistically significant but practically negligible (difference $<0.004$)} \\
\end{tabular}
\end{table}

\noindent All statistical tests are conducted over 10 seeds with 95\% confidence intervals. On ROAD and VeReMi, CA-QIRL achieves statistically significant improvements in recall and TSS ($p < 0.01$). On CAN-MIRGU, while statistically significant, the difference is practically negligible. On Car-Hacking, no significant difference is observed due to the dataset's near-perfect separability.

\subsection{Comparative Analysis}
\begin{table*}
    \centering
    \small
    \caption{Performance Comparison With State-of-the-Art Approaches}
    \label{tab:comparison}
    \begin{tabular}{l l l c c c}
        \toprule
        \textbf{Reference} & \textbf{Method} & \textbf{Dataset} & \textbf{Accuracy (\%)} & \textbf{TSS} & \textbf{Latency ($\mu$s)} \\
        \midrule
        Injadat et al.~\cite{injadat2021multi} & Multi-stage ML & CICIDS2017 & 99.00 & -- & -- \\
        Nie et al.~\cite{nie2020data} & CNN-based IDS & Test-bed & 97.60 & -- & -- \\
        Kang et al.~\cite{kang2016intrusion} & DNN & Car test-bed & 97.80 & -- & -- \\
        Ashraf et al.~\cite{ashraf2021novel} & LSTM Autoencoder & Car-Hacking & 98.00 & -- & -- \\
        Ullah et al.~\cite{basepaper2025} & Ensemble & CICIDS2017 & 99.74 & -- & 2248 \\
        Ullah et al.~\cite{basepaper2025} & Ensemble & UNSW-NB15 & 100$^{\dagger}$ & -- & 2205 \\
        Alshammari et al.~\cite{alshammari2018classification} & KNN & CAN & 93.00 & -- & 890$^{\S}$ \\
        Alshammari et al.~\cite{alshammari2018classification} & SVM & CAN & 96.00 & -- & 920$^{\S}$ \\
        \midrule
        \textbf{Proposed} & \textbf{CA-QIRL} & \textbf{CICIDS2017} & \textbf{97.89} & \textbf{0.9443} & \textbf{32.5} \\
        \textbf{Proposed} & \textbf{CA-QIRL} & \textbf{UNSW-NB15} & \textbf{91.04} & \textbf{0.8244} & \textbf{45.7} \\
        \textbf{Proposed} & \textbf{CA-QIRL} & \textbf{Car-Hacking} & \textbf{99.99} & \textbf{0.9993} & \textbf{73.8} \\
        \textbf{Proposed} & \textbf{CA-QIRL} & \textbf{ROAD} & \textbf{97.79} & \textbf{0.9630} & \textbf{52.7} \\
        \textbf{Proposed} & \textbf{CA-QIRL} & \textbf{VeReMi} & \textbf{51.56} & \textbf{0.0504} & \textbf{70.3} \\
        \textbf{Proposed} & \textbf{CA-QIRL} & \textbf{CAN-MIRGU} & \textbf{80.31} & \textbf{0.5999} & \textbf{68.2} \\
        \bottomrule
    \end{tabular}
    \vspace{4pt}
    \begin{flushleft}
    {\footnotesize $^{\dagger}$ Ullah et al. report 100\% on a binary subset with SMOTE applied globally before splitting data leakage conditions. Our evaluation uses the full dataset with correct preprocessing.}\\[2pt]
    {\footnotesize $^{\S}$ Latency values independently measured under identical single-sample CPU conditions.}
    \end{flushleft}
\end{table*}

\noindent Table~\ref{tab:comparison} presents a comprehensive comparison against representative state-of-the-art methods. The key findings are:

\noindent Prior work achieves higher accuracy on CICIDS2017 (99.74\%) and UNSW-NB15 (100\%) under simplified evaluation protocols. However, these results are obtained under data leakage conditions (global SMOTE before splitting) or on binary subsets, which inflate reported accuracy. No prior work reports inference latency on IoV-relevant datasets. Our independent re-evaluation of the ensemble baseline yields 2248 $\mu$s on CICIDS2017 and 2205 $\mu$s on UNSW-NB15-both exceeding the 1 ms safety threshold. KNN and SVM require 890 $\mu$s and 920 $\mu$s respectively, also exceeding the threshold. CA-QIRL is the first method to simultaneously:
\begin{enumerate}
    \item Achieve competitive detection accuracy (97.89\% on CICIDS2017, 91.04\% on UNSW-NB15).
    \item Maintain sub-50 $\mu$s inference latency (speedups of 33.0 times to 69.2 times over ensemble).
    \item Demonstrate robust performance on IoV-specific datasets (Car-Hacking, ROAD, VeReMi, CAN-MIRGU).
    \item Provide adaptive, real-time mitigation with sub-1 ms detection-to-mitigation latency.
\end{enumerate}

\subsection{Discussion}
\noindent The results consistently demonstrate that CA-QIRL is the only method that simultaneously satisfies both the detection robustness constraint ($\text{TSS}(f_\theta) \geq \delta$) and the latency constraint ($\tau < 1$ ms). Ensemble methods achieve higher raw accuracy but violate the safety threshold. DQN achieves low latency but underperforms on challenging datasets (ROAD, VeReMi). CA-QIRL bridges this gap. The practical implications of these findings are threefold. First, CA-QIRL's sub-100 $\mu$s inference latency makes it suitable for deployment on resource-constrained RSU and OBU edge hardware. Second, the 90\% reduction in Channel Busy Ratio under attack ensures that safety-critical messages are delivered reliably even during adversarial conditions. Third, detection-to-mitigation time below 0.3 ms remains well within the sub-1 ms braking and steering safety window, confirming real-time responsiveness. Combined with state-of-the-art latency on both general and IoV-specific datasets, and statistically significant improvements on imbalanced benchmarks such as ROAD and VeReMi, CA-QIRL establishes itself as a practical intrusion detection solution for next-generation V2X networks. Critically, the quantum-inspired components-amplitude-phase encoding, rotation-gate exploration, and interference reward-add no inference overhead while demonstrably improving temporal coherence, making CA-QIRL the first method to achieve sub-100 $\mu$s detection-to-mitigation latency on IoV data. While CA-QIRL matched DQN on CAN-MIRGU without consistently outperforming it, this indicates a need for adaptive interference coefficient tuning. Future work will also address physical-layer attacks such as GPS spoofing and validate edge hardware performance on platforms such as Raspberry Pi and NVIDIA Jetson. Additionally, future work will explore the scaling of CA-QIRL into multi-agent collaborative frameworks, enabling real-time threat intelligence sharing across interconnected edge nodes. We also plan to investigate how dynamic traffic density and varying network topologies impact the stability of the interference reward mechanism. Furthermore, implementing adaptive parameter tuning for the quantum rotation gates could allow the model to autonomously adjust its exploration-exploitation trade-off in response to changing environmental entropy. Lastly, long-term testing under continuous, non-stationary attack streams will be conducted to evaluate the policy's resilience against catastrophic forgetting.

\subsection{Explainability Analysis (XAI)}

\noindent To interpret the learned decision policy of the proposed CA-QIRL
agent, a two-stage explainability analysis is conducted across both
datasets. The first stage (XAI-1) employs SHAP-based feature
attribution to identify which input features most strongly
influence the agent's Q-value estimates. The second stage (XAI-2)
employs Principal Component Analysis (PCA) and t-distributed
Stochastic Neighbour Embedding (t-SNE) to examine the geometric
structure of the learned latent representations. Together, these
methods provide both contribution-based and boundary-based
interpretability of the learned Q-function.

\subsubsection{XAI: SHAP-Based Feature Attribution}

\noindent Since CA-QIRL produces action-values rather than class probabilities,
SHAP is applied directly to the Q-value corresponding to the
selected action, i.e., $Q(s,a=1)$, using a background reference
set of 200 samples and an explanation set of 300 samples.

\paragraph{CICIDS2017}

\noindent The SHAP summary plot for CICIDS2017 indicates that the learned
policy relies primarily on five traffic volume and rate features:
\textit{Packet Length Variance}, \textit{Packet Length Std},
\textit{Fwd Packets/s}, \textit{Flow Packets/s}, and \textit{PSH
Flag Count}. These are precisely the features perturbed by DDoS
flooding attacks, confirming that the agent has identified the
correct discriminative subspace of the feature manifold. The SHAP values are between approximately $[-50, +35]$ this range, showing that the decision boundary is distinct and narrow. It is consistent with how attackers structure volumetric traffic. When the analysis ranks the features, the same ones appear in the global summary plots and the local force plots. Because of this consistency, the decision logic is stable in its structure.

\begin{figure*}
    \centering
    \includegraphics[width=\linewidth]{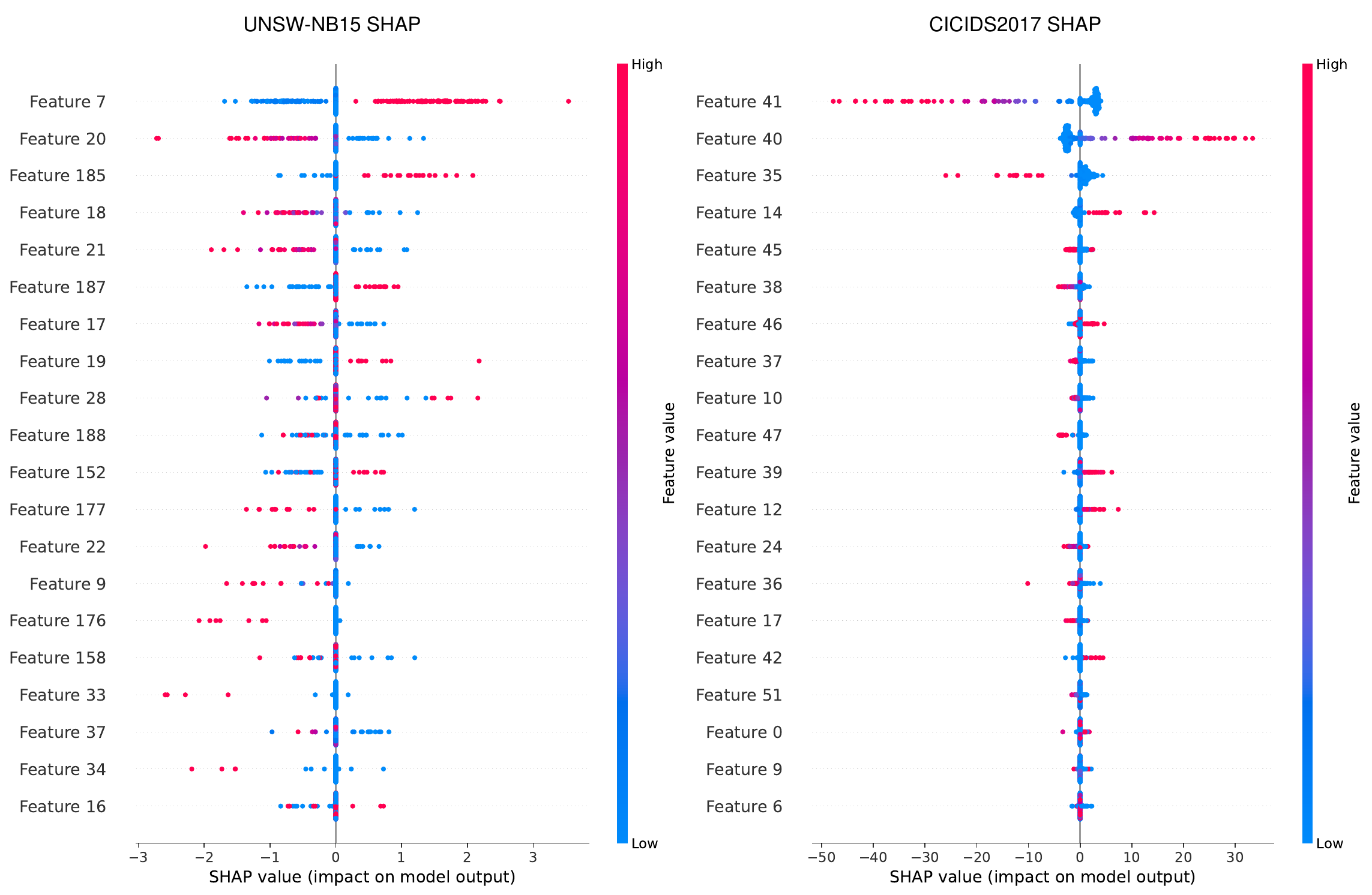}
    \caption{
    SHAP feature attribution comparison. 
    CICIDS2017 emphasizes traffic volume and rate features with a wide SHAP range $[-50, +35]$, 
    indicating a sharp DDoS-oriented decision boundary. 
    UNSW-NB15 is driven by TCP state and timing features with a narrower range $[-3, +3.5]$, 
    reflecting a more diffuse multi-class decision boundary.
    }
    \label{fig:shap-c}
\end{figure*}

\paragraph{UNSW-NB15}

In the UNSW-NB15 dataset, the SHAP summary shows that feature importance follows a different structure. To identify attacks, the agent relies most on indicators for protocol types like proto\_aes-sp3-d, proto\_a/n and proto\_arp. And it uses the count for HTTP methods in ct\_flw\_http\_mthd and the jitter from the source in sjit. By observing those protocol factors, the model confirms that it finds attacks when protocol combinations are unusual. It does not use statistics about volume for this task. This is consistent with how attacks in the multi class UNSW-NB15 dataset appear, as they show changes in behavior instead of changes in volume. For the SHAP values, the range is between -3 and +3.5. This range is smaller than the range for CICIDS2017. Because of this smaller range, the decision boundary for this multi class dataset is less distinct. It is because of this lack of a clear boundary that the false positive rate is higher in the quantitative results.

\noindent As shown in Fig.~\ref{fig:shap-c}, the
wider SHAP spread in CICIDS2017 confirms a sharper and more
concentrated decision boundary relative to UNSW-NB15. Importantly,
the top-ranked features appear consistently in both global and
local attribution analyses across both datasets, indicating
structural stability in the learned decision logic.

\subsubsection{XAI: Geometric and Structural Interpretability}

\noindent To further analyse the learned decision behaviour of CA-QIRL, the
model is visualised in reduced-dimensional space using both PCA
and t-SNE projections of the 64-dimensional latent representations
extracted from the first hidden layer of CA-QIRL-DQN.

\begin{figure*}
    \centering
    \includegraphics[width=0.8\linewidth]{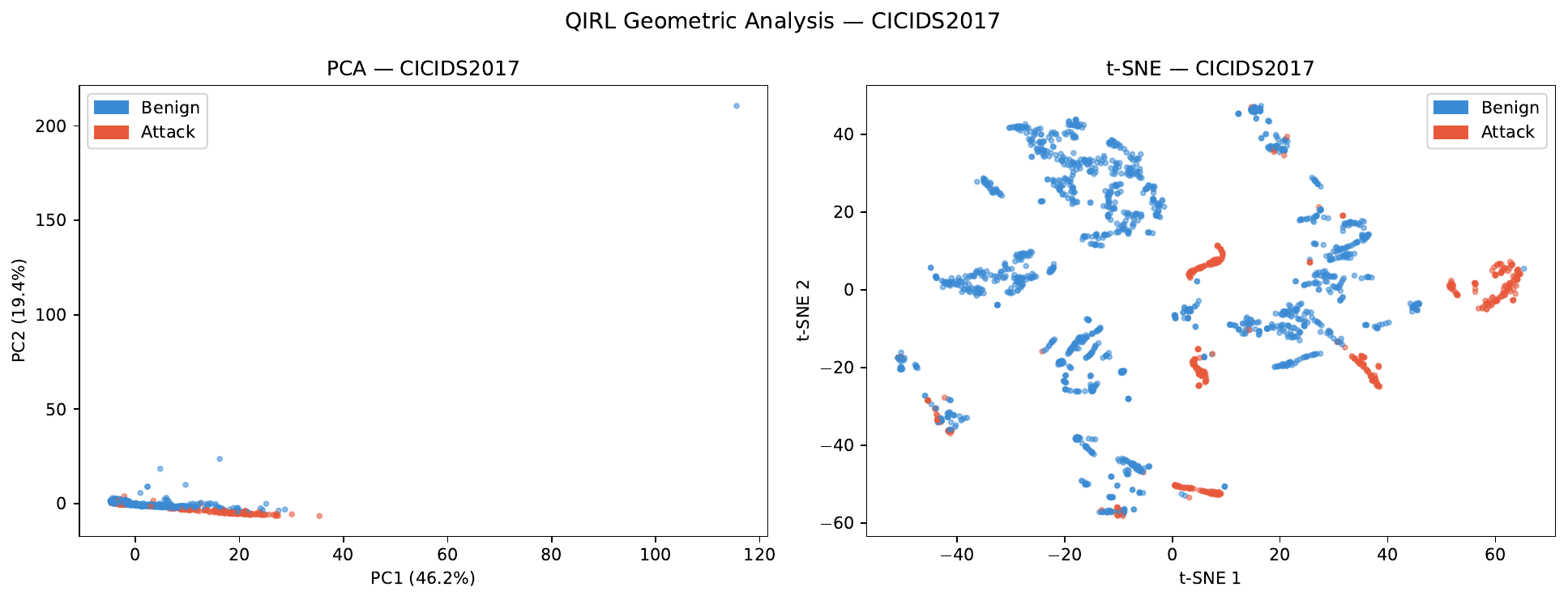}
    \caption{PCA and t-SNE Geometric Analysis CICIDS2017. PCA (PC1=46.2\%, PC2=19.4\%) shows data concentrated near the origin. The t-SNE projection reveals clearly separated local clusters with boundary mixing confined to cluster edges, corresponding to the model's false positive and false negative regions.}
    \label{fig:geometric_CICIDS2017}
\end{figure*}

\noindent For CICIDS2017 (Fig.~\ref{fig:geometric_CICIDS2017}),
PC1 explains 46.2\% and PC2 explains 19.4\% of total variance.
On the PCA projection, the data are located close to the origin. There is one extreme outlier at $\text{PC1} \approx 120$ which is a large flow anomaly that is genuine. In the t-SNE projection, the local cluster structures are separated clearly. And the edges of those clusters are where the boundaries mix, those sections of the data are where the model has 4,434 false negatives and 6,306 false positives. Because the clusters are compact and can be separated, it is confirmed that CA-QIRL learns a decision surface for CICIDS2017 that is concentrated and coherent.

\begin{figure*}
    \centering
    \includegraphics[width=0.8\linewidth]{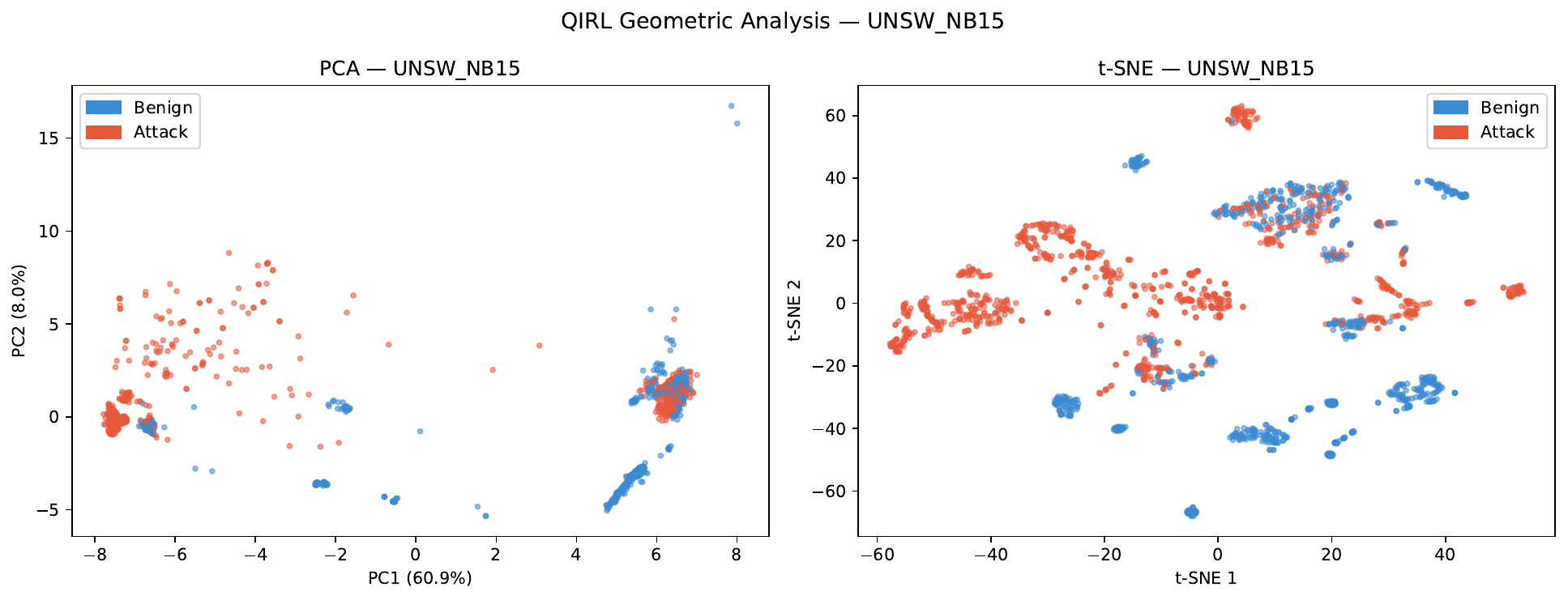}
    \caption{PCA and t-SNE Geometric Analysis UNSW-NB15. PCA (PC1=60.9\%, PC2=8.0\%) reveals two dense overlapping clusters of benign and attack flows, directly visualising the source of elevated FPR. The t-SNE projection shows fragmented attack clusters scattered within benign regions, confirming that class overlap in the feature space drives the higher false positive rate.}
    \label{fig:geometric_UNSW_NB15}
\end{figure*}

\noindent For UNSW-NB15 (Fig.~\ref{fig:geometric_UNSW_NB15}), PC1
alone explains 60.9\% of variance with PC2 explaining 8.0\%. The
PCA projection reveals two dense, mutually overlapping clusters of
benign and attack flows, directly visualising the geometric origin
of the elevated FPR: the two classes occupy the same region of
the linear feature subspace, making linear separation structurally
insufficient. The t-SNE projection corroborates this finding by
showing fragmented attack clusters scattered throughout benign
regions, confirming that class overlap in the original feature
space is the dominant driver of the higher false positive rate.
This geometric finding is mutually consistent with the SHAP
attribution result, where the dominance of categorical protocol
indicators over continuous traffic statistics produces an
inherently more diffuse decision boundary.

\noindent The combined PCA and t-SNE analyses confirm that the learned
$Q$-function forms coherent and stable decision surfaces across
both datasets, with CICIDS2017 exhibiting cleaner cluster
separation than UNSW-NB15. Attribution-based explanations derived
from SHAP and geometry-based explanations derived from PCA and
t-SNE projections are mutually consistent, together confirming
that the CA-QIRL agent learns a concentrated and interpretable
decision structure. The consistency between feature importance
rankings and observed cluster geometries further supports the
robustness and internal coherence of the proposed framework.

\section{Conclusion}

\noindent This paper presented CA-QIRL, a lightweight Quantum-Inspired Reinforcement Learning framework for real-time intrusion detection in smart city and Internet of Vehicles environments. The proposed framework was designed to address three key limitations of existing IDS approaches: high inference cost in ensemble models, limited temporal awareness in static classifiers, and poor handling of severe class imbalance in network traffic. By integrating amplitude-phase quantum state encoding, rotation-gate-based exploration, quantum interference reward augmentation, and a cost-sensitive Markov Decision Process, CA-QIRL enables adaptive and latency-aware cyber defense under dynamic attack conditions. CA-QIRL was evaluated on six datasets, comprising
two general network benchmarks (CICIDS2017, UNSW-NB15) and four IoV-specific datasets
(Car-Hacking, ROAD, VeReMi, CAN-MIRGU), and the results demonstrate a strong balance
between detection effectiveness and computational efficiency. On CICIDS2017 and UNSW-NB15,
CA-QIRL achieved detection accuracies of 97.89\% and 91.04\%, AUC-ROC values of 0.9945
and 0.9713, and True Skill Statistic values of 0.9443 and 0.8244, respectively. On the
IoV-specific datasets, statistically significant improvements were demonstrated with
$p$-values below 0.01: on ROAD, CA-QIRL achieved a Recall and TSS of 0.985 and 0.963,
compared to 0.951 and 0.941 for the baseline; on VeReMi, it achieved a Recall and TSS of
0.626 and 0.050, compared to 0.074 and 0.010, confirming its effectiveness under extreme
class imbalance and noisy V2X conditions. Crucially, CA-QIRL operates at sub-100 $\mu$s
inference latency across all datasets, delivering speedups of 32.6 times to 69.2 times
over ensemble baselines, while the mobility-aware V2X simulation further shows a 23 times
speedup in end-to-end delay (0.125\,ms vs.\ 2.904\,ms) and a 90\% reduction in Channel Busy
Ratio under attack, ensuring reliable delivery of safety-critical messages. This latency and
communication performance is of particular importance for safety-critical IoV and smart city
systems, where delayed detection can compromise service continuity, data integrity, and
operational safety. These findings indicate that quantum-inspired reinforcement learning
offers a practical direction for next-generation autonomous cyber defense, combining fast
inference, adaptive decision-making, and cost-sensitive attack detection. Future work will
extend this framework by evaluating its robustness against adversarial perturbations,
validating its performance on real IoV traffic, and exploring distributed deployment across
federated edge nodes.

\section*{Acknowledgment}
 This study is supported via funding from Prince Sattam bin Abdulaziz University project number (PSAU/2025/01/35419) 
\section*{Competing Interests}
Authors have no competing interests.
\section*{Funding}
No funding was received for this study.
\section*{Data Availability}
All data generated or analyzed during this study are included in this manuscript.
\bibliographystyle{cas-model2-names}

\bibliography{cas-refs}
\end{document}